\documentclass[conference]{IEEEtran}

\usepackage{graphicx} 
\usepackage{lipsum} 
\usepackage{fancyhdr} 
\usepackage{amsmath}
\usepackage[english]{babel}
\usepackage[utf8]{inputenc}
\usepackage[T1]{fontenc}
\usepackage{lmodern}
\usepackage{amsmath,amssymb,amsfonts}
\usepackage{amsthm}
\usepackage[draft]{hyperref}
\usepackage{enumitem}
\usepackage{graphicx}
\hypersetup{hidelinks}
\usepackage{float}

\title{A Preliminary Model for Managing Technical Debt in an Agile Environment}

\author{
\IEEEauthorblockN{Dr. Pedro E. Colla}
\IEEEauthorblockA{
UADER-FCyT\\
E-Mail: pedroernesto.colla@ufasta.edu.ar, colla.pedro@uader.edu.ar
}
}

\begin{document}

\maketitle

\pagestyle{fancy}
\fancyhf{} 

\lfoot{Performed under grant UADER FCyT Project PI-B 230/24} 
\rfoot{\thepage}        

\newtheorem{theorem}{Theorem}
\begin{abstract}
This paper presents a preliminary model for managing involuntary technical debt in agile environments by formulating, in an integrated way, the dynamics among backlog, debt, velocity, and economic value. The work distinguishes initiated but unfinished functional debt from a simple \textit{defect backlog} and from rework, interprets productivity degradation as technical-debt interest, and derives the naive maximum-remediation policy in order to show its limitations against an intertemporal value-based decision. On this basis, a dynamic policy $u_k$ is proposed to balance new development and remediation; a decreasing marginal-value structure is incorporated; and the model is extended to discrete, inhomogeneous items. Exploratory validation through sensitivity analysis and Monte Carlo simulation shows behavior consistent with the economic intuition of the model. Finally, the limits of the formulation are made explicit: its macroscopic nature, its dependence on organizationally stable parameters, its assumption of intertemporal rationality, and its requirement of weak coupling among stories.
\end{abstract}

\section{Introduction}

The notion of \textbf{technical debt} has become a central construct in modern software engineering, particularly in the context of agile methods such as Scrum, where incremental delivery and the pressure to generate early value introduce structural tensions among speed, quality, and sustainability. The term was originally coined by Ward Cunningham \cite{cunningham1992wycash} to describe the deferred cost of suboptimal design decisions, conceptualizing debt as a conscious commitment that allows development to be accelerated in the short term at the expense of future costs.

Since then, the literature has evolved toward a broader and more formal characterization of technical debt. Works such as those by Steve McConnell \cite{mcconnell} and Philippe Kruchten et al. \cite{kruchten2012technical, kruchten2019} distinguish among different types of debt (code, design, architecture, testing, documentation), emphasizing its multifaceted nature. Nicolas Brown et al. \cite{brown2010managing}, in turn, introduce a more refined taxonomy that differentiates between \textbf{deliberate} and \textbf{inadvertent} debt, and between \textbf{prudent} and \textbf{reckless} debt, providing a key conceptual framework for the analysis of technical decisions under uncertainty.

In agile development, several authors have noted that technical debt emerges as an almost inevitable consequence of the iterative model. Mary Poppendieck and Tom Poppendieck \cite{poppendieck2003lean} had already warned about the risks of accumulating technical waste in systems subject to fast delivery cycles. More recently, Ipek Ozkaya, Philippe Kruchten , and Robert L. Nord \cite{kruchten2019} have examined the relationship between technical debt and architectural decisions, stressing the need to manage debt explicitly as part of the agile planning process.

\subsection{Technical debt management: existing approaches}

Approaches proposed for technical debt management may be grouped into three broad categories:

\begin{enumerate}
    \item \textbf{Qualitative and heuristic approaches} \\
    Based on engineering practices such as code reviews, continuous refactoring, and Definition of Done. Authors such as Martin Fowler \cite{fowler1999refactoring} emphasize systematic refactoring as a mechanism for controlling emerging debt.

    \item \textbf{Metric-based approaches} \\
    Based on indicators such as cyclomatic complexity, test coverage, or maintainability index. Capers Jones \cite{jones2010bestpractices} and Tom Mens \& Serge Demeyer \cite{mens2008softwareevolution} have contributed to formalizing the measurement of structural quality as a proxy for debt.

    \item \textbf{Economic approaches} \\
    These regard debt as a financial phenomenon involving interest and principal costs. Yannis Smaragdakis and Nicolas Brown, together with Ipek Ozkaya \cite{brown2010managing}, have promoted models in which debt is managed in terms of economic trade-offs and business value.
\end{enumerate}

More recently, quantitative models integrating system dynamics and economic theory have been proposed, for example, by Hakan Erdogmus and Nari Sangal \cite{erdogmus2016technical}), although these models usually assume aggregate structures and do not fully capture the granularity of sprint-level decisions.

\subsection{Involuntary technical debt}

A less explored aspect in the literature is \textbf{involuntary technical debt}, understood as debt that emerges not from a strategic decision but as a consequence of defects, implementation errors, or limitations in the development process. This form of debt is closely related to the quality of the testing process and the effectiveness of early assurance practices.

Authors such as Barry Boehm \cite{boehm1981softwareeconomics} had already pointed out the importance of defects as generators of cumulative costs throughout the life cycle. Along these lines, defect-flow models (defect inflow/outflow) have been used to characterize the dynamics of technical-problem accumulation.

More recently, Bavani \cite{bavani2012testingdebt} and Forrest Shull \cite{shull2011perfectionists} have explored the relationship between testing practices and technical debt, while Iftekhar Ahmed et al. \cite{ahmed2015technicaldebt} have studied the correlation between residual defects and structural debt.

However, most of these approaches:
\begin{itemize}
    \item treat defects as discrete events without modeling their cumulative economic impact,
    \item do not explicitly integrate functional backlog and debt into the same dynamic framework,
    \item and lack a formulation that makes it possible to optimize intra-sprint decisions under uncertainty.
\end{itemize}

\subsection{Relationship among \textit{defect backlog}, rework, and involuntary technical debt}
\label{subsec:defect-backlog-rework-deuda-involuntaria}

To avoid ontological ambiguities, this paper explicitly distinguishes among three related but non-equivalent concepts: \textit{defect backlog}, rework, and involuntary technical debt. All three may originate in residual defects or limitations of the quality-assurance process, but they represent different objects within the model.

The \textit{defect backlog} is understood as the observable set of known, recorded defects pending correction. It is therefore a queue of discrete events: each defect has a description, a severity, a detection date, a state, and eventually a priority. Its natural management unit is the individual defect. In contrast, involuntary technical debt is not necessarily identified with all known defects and is not exhausted by them. It represents the stock of pending technical effort embedded in the system when a functionality was initiated, partially or fully integrated into the product, but did not satisfactorily complete the expected cycle of construction, verification, validation, and stabilization.

Rework, in turn, is not a stock but a flow of consumed capacity. In the notation of the model, $R_k$ represents story points applied in sprint $k$ to remediation. This effort may be used to correct specific defects, stabilize components, refactor a defective solution, complete omitted tests, or close residual technical work. For this reason, rework is the action that reduces accumulated debt, totally or partially; it is not the debt itself. This distinction justifies modeling debt as a state variable $D_k$, while remediation appears as a flow variable determined by the capacity-allocation decision $u_k$.

Involuntary technical debt is therefore modeled as ``functionality initiated but not completed'' because the defect that gives rise to it is not interpreted only as an isolated anomaly, but as evidence that a unit of functional value passed through the development process without reaching a sufficiently stable completion condition. In an agile environment, a story may have been designed, coded, integrated, partially demonstrated, or even released and still require subsequent work to reach the expected quality level. That situation differs from an ordinary functional backlog: it is not functionality not yet initiated, but functionality that has already consumed capacity, captured a fraction of value, and left a technical residue that affects future velocity.

This formulation also explains the introduction of $\theta_k$ as the average fraction of value already captured by debt items. If a unit of debt comes from partially completed functionality, not all of its value remains pending. The fraction $1-\theta_k$ represents the residual value recoverable through remediation. Consequently, involuntary debt occupies an intermediate position between pure functional backlog and punctual defect: it preserves a relationship with already-prioritized business value, but technically manifests itself as pending stabilization, correction, or completion work.

The conceptual separation may be summarized as follows:
\begin{itemize}
    \item the \textit{defect backlog} records known defects as discrete units of observation and management;
    \item rework measures the flow of effort consumed to correct, stabilize, or complete previously initiated work;
    \item involuntary technical debt measures the aggregate stock of residual technical effort embedded in the system as a consequence of defects, omissions, or insufficient early containment of errors.
\end{itemize}

Under this interpretation, a defect may be the observable signal that increases the \textit{defect backlog}; rework is the capacity allocated to resolve it; and involuntary technical debt is the persistent state that captures the accumulated effect of those insufficiencies on the future capacity of the system. Therefore, the model does not replace operational defect management or rework measurement, but abstracts their dynamic impact into a state variable expressed in story points. This abstraction makes it possible to integrate the phenomenon into the same decision structure that governs capacity allocation between new functionality and remediation.

Finally, this delimitation avoids two problematic reductions. First, it avoids reducing involuntary debt to a simple list of defects, because many defects have systemic effects, technical dependencies, and stabilization costs that exceed punctual correction. Second, it avoids treating all rework as debt, because there may be local, minor, or immediately absorbed rework that does not generate a persistent stock and does not degrade future velocity. In the proposed model, only that portion of residual work that persists between sprints, competes for capacity with the functional backlog, and affects future productivity through the degradation represented by $D_k$ is considered involuntary technical debt.

\subsection{Contribution of this work}

This paper proposes a formal model that integrates:

\begin{itemize}
    \item the simultaneous evolution of the \textbf{functional backlog} and \textbf{technical debt},
    \item sprint capacity allocation through a control variable $u_k$,
    \item the dynamics of debt generation, both deliberate and involuntary,
    \item and the economic valuation of the system through discounted-value functions.
\end{itemize}

Unlike previous work, this model:

\begin{enumerate}
    \item \textbf{Explicitly integrates involuntary debt} \\
    By modeling debt generation as a function of residual defects and process-quality parameters.

    \item \textbf{Introduces a discrete optimal-control formulation} \\
    Where the decision to allocate effort between development and remediation is optimized in each sprint.

    \item \textbf{Unifies technical and economic perspectives} \\
    By linking quality metrics with present-value impact, which makes it possible to analyze trade-offs in quantitative terms.

    \item \textbf{Enables simulation under uncertainty (Monte Carlo)} \\
    By incorporating probability distributions over key parameters, enabling sensitivity analysis and robust policy evaluation.
\end{enumerate}

Although partial precedents exist in economic models of technical debt (for example, in the work of Ipek Ozkaya et al.), no model has been identified in the literature that simultaneously combines:
\begin{itemize}
    \item backlog and debt dynamics,
    \item intra-sprint control,
    \item defect-based involuntary debt,
    \item and integrated economic valuation.
\end{itemize}

In this sense, this work contributes to closing a gap between the conceptual literature and the need for operational analytical tools for decision-making in agile environments.

\section{Initial Problem Statement}

Let $k=0,1,\dots,H-1$ be the sprint index, where $H \in \mathbb{N}$ is the total analysis horizon.

We define:

\begin{equation}
B_k \in \mathbb{R}_{\ge 0}
\tag{1}
\end{equation}

as the pure backlog remaining at the beginning of sprint $k$, with $B_0$ being the total backlog at the beginning of the project, all measured in effort units as the equivalent amount of new functionality not yet developed. As is natural in agile methods, during development the project team may decide to defer a given function to the future; this may be understood as a reprioritization of the backlog based on technical or operational needs. It's common in the literature to regard these changes in priority as a form of \textit{voluntary technical debt}; however, when they are based on an objective criterion, they may be assimilated to backlog reprioritization and imply knowledge of the variation in value.
A different condition arises when technical debt consists of functions that were duly prioritized and whose development was attempted, but which, due to problems of different kinds, did not successfully complete the development cycle. In this case, the problems may be treated as \textit{defects} which, as such, can be solved during the rest of the project. This group is sometimes referred to as \textit{involuntary technical debt} and is denoted by $D_{k}$ such that it satisfies:
\begin{equation}
D_k \in \mathbb{R}_{\ge 0}
\tag{2}
\end{equation}
The stock of remaining functional debt at the beginning of sprint $k$, measured as the equivalent amount of functionality already initiated but not fully realized; at the same time, the development rate, in units of functional deployment per sprint, is characterized as:
\begin{equation}
V_0>0
\tag{3}
\end{equation}
as the \textit{base velocity} of the system in the absence of technical debt. It is expected that technical debt requires more effort to resolve, since it may require additional activities beyond pure development, such as research, impact analysis, or others. To account for this, we propose a degradation function such that the velocity in a given sprint $k$ is represented by $V_k$, whose expression is:
\begin{equation}
V_k=\frac{V_0}{1+\gamma D_k}
\label{eq:5}
\tag{4}
\end{equation}
with
\begin{equation}
\gamma \ge 0
\tag{5}
\end{equation}
being the parameter that measures how much debt deteriorates the effective velocity of the system as a result of the technical debt that must be managed.

This loss of velocity can be interpreted as a form of \textit{technical-debt interest}: debt not only represents pending remediation work, but also increases the future cost of producing new functionalities and modifying the system. In the literature, technical-debt interest has been associated both with greater maintenance and evolution effort and with accumulated productivity loss while debt remains unremediated \cite{kamei2016interest,ampatzoglou2015interest,martini2017interest}. In the present model, this effect is summarized macroscopically through the parameter $\gamma$, which translates the aggregate stock $D_k$ into degradation of the effective velocity $V_k$.

\subsection{Budgetary fit and fundable horizon}

Before continuing to study the problem of dynamic allocation between new backlog and remediation, it is useful to make explicit the budgetary fit of the problem. In this subsection, technical-debt effects on productivity or value are not yet incorporated; the goal is only to link budget, team cost, and the time horizon available to execute the project.
Let:
\begin{equation}
B > 0
\tag{6}
\end{equation}
where B is the total budget available for the project, expressed in monetary units;
\begin{equation}
p \in \mathbb{N}
\tag{7}
\end{equation}
the number of engineers assigned to the project, assumed constant, and
\begin{equation}
CPE > 0
\tag{8}
\end{equation}
the average cost per engineer per month during the project; at the same time:
\begin{equation}
t_{SP} > 0
\tag{9}
\end{equation}
the duration of each sprint, expressed as fractions of a time unit dimensionally consistent with the rest of the parameters and assumed uniform during the project.
Under these simplified definitions, the monetary cost of one sprint is:
\begin{equation}
C_{sprint} = p \cdot CPE \cdot t_{SP}
\tag{10}
\end{equation}
The maximum number of sprints fundable by the budget ($K_{B}$) is:

\begin{equation}
C_0 = K_{B}  \cdot C_{sprint}
\tag{11}
\end{equation}

\begin{equation}
K_B = \left\lfloor \frac{B}{C_{Sprint}} \right\rfloor
=
\left\lfloor \frac{B}{p \cdot CPE \cdot t_{SP}} \right\rfloor
\tag{12}
\end{equation}

The effective time horizon of the project, expressed in months, is given by:
\begin{equation}
T_B = K_B \cdot t_{SP}
\tag{13}
\end{equation}
and, expressed directly in sprint units, by:
\begin{equation}
H = K_B
\tag{14}
\end{equation}
Thus, even before considering the dynamics of technical debt, the problem is naturally bounded by a \textbf{finite} horizon imposed by the budget constraint. Consequently, any subsequent intertemporal formulation must be solved over $k=0,\dots,K_B-1$.

\section{Capacity Applied to New Functionality and Remediation}

As the project evolves, it is necessary to balance what proportion of effort is devoted to implementing new functionality and what proportion is devoted to resolving \textit{involuntary} technical debt.
Let:
\begin{equation}
N_k \in \mathbb{R}_{\ge 0}
\tag{15}
\end{equation}
be the part of sprint $k$ capacity applied to developing new functionalities, and let:
\begin{equation}
R_k \in \mathbb{R}_{\ge 0}
\tag{16}
\end{equation}
be the part applied to debt remediation.
The aggregate sprint capacity constraint is:
\begin{equation}
N_k + R_k = V_k
\tag{17}
\end{equation}
The evolution of the pure backlog is given by:
\begin{equation}
B_{k+1}=\max(0,\;B_k-N_k)
\tag{18}
\end{equation}
Development activity generates defects from an injection process that the development methodology addresses but will almost certainly not eliminate completely. To take this situation into account, we introduce two factors:

\begin{enumerate}[label=\arabic*.]
\item new development generates debt at rate $\alpha$,
\item remediation eliminates debt with net effectiveness $1-\beta$.
\end{enumerate}

We then define:
\begin{equation}
\alpha \in [0,1)
\tag{19}
\end{equation}
to account for the fact that building one unit of new work will introduce defects, and
\begin{equation}
\beta \in [0,1)
\tag{20}
\end{equation}
which also captures the fact that debt reduction will introduce defects during remediation, that is, the inefficient fraction of remediation effort.
With this, debt evolves as:
\begin{equation}
D_{k+1}=\max\left(0,\;D_k+\alpha N_k-(1-\beta)R_k\right)
\tag{21}
\end{equation}
Previous equations constitute the aggregate dynamic core of the problem.
The factors $\alpha$ and $\beta$ can be estimated for a given organization from historical data by means of linear-regression techniques. Experience shows that, as organizations progress in process maturity, they reach stability and even statistical capability in this type of metric.
The use of these parameters, as well as the parameter $\gamma$ introduced to represent the sensitivity of velocity to technical debt, presupposes that the organization has sufficient process maturity to estimate relatively stable historical metrics. This assumption, together with other validity conditions of the model, is made explicit later in the limitations section.

\section{Impact of Technical Debt on Horizon and Budget}

This section extends the base sprint-planning model by explicitly incorporating the effect of technical debt through the previously defined parameters $\alpha$, $\beta$, and $\gamma$.

\subsection{Macroscopic scope of the approximation}

Before introducing the velocity-degradation dynamics, it is useful to specify the scope of the adopted representation. In real systems, the impact of technical debt on productivity is rarely uniform, linear, or homogeneously distributed over the whole product. Frequently, debt concentrates in specific components, critical modules, unstable interfaces, areas with low test coverage, or regions of the code that concentrate a disproportionate share of changes and defects. These concentration points often operate as technical \textit{hotspots} and may produce nonlinear effects, degradation thresholds, operational discontinuities, or local blockages that do not appear in proportion to the aggregate volume of debt.

Therefore, the use of an aggregate variable $D_k$ and a sensitivity parameter $\gamma$ should not be interpreted as a claim that every unit of debt has the same marginal impact on team velocity. The expression that links $D_k$ with the effective velocity $V_k$ constitutes a macroscopic approximation of the average systemic effect of technical debt on delivery capacity. In this sense, $D_k$ summarizes a heterogeneous technical stock, while $\gamma$ captures an aggregate sensitivity that may depend on architecture, modularity, debt distribution, the criticality of affected components, and the team's ability to isolate or contain its effects.

This clarification is important because the continuous model developed below operates on stocks of effort and not on individual technical items. Its purpose is to represent the economic relationship among backlog, debt, velocity, and remediation at the sprint-planning level, not to replace detailed analysis of architecture, component dependencies, or defect localization. The discrete extension presented later partially recognizes the indivisibility and heterogeneity of debt items, but even there the model retains an aggregate reading oriented toward economic decision-making.

Consequently, the model should be understood as a first-order formulation. When debt is relatively distributed, or when local effects can be averaged at sprint level, the aggregate approximation may be appropriate for comparing capacity-allocation policies between new functionality and remediation. In contrast, when dominant \textit{hotspots}, severe technical thresholds, or critical architectural dependencies exist, the variable $D_k$ should be complemented with localization factors, criticality weights, or specific constraints representing the nonlinear and localized character of the impact of debt. These extensions do not contradict the basic formulation; rather, they delimit its validity domain and indicate how it could be empirically enriched in subsequent applications.

\subsection{Dynamic model with technical debt}

Starting from the expression that defines $V_k$ = f($V_0$,$D_k$,$\gamma$), the capacity of a given sprint $k$ can be distributed between new development ($N_k$) and remediation ($R_k$):
\begin{equation}
N_k = (1 - u_k) V_k
\tag{22}
\end{equation}
\begin{equation}
R_k = u_k V_k
\tag{23}
\end{equation}
where $u_k \in [0,1]$ represents the fraction of effort dedicated to remediation. In that case, the technical-debt dynamics are modeled as:
\begin{equation}
D_{k+1} = \max\left(0,\; D_k + \alpha N_k - (1 - \beta) R_k \right)
\tag{24}
\end{equation}
Let $B_k$ be the remaining functional backlog at the beginning of sprint $k$. Its evolution is given by:
\begin{equation}
B_{k+1} = \max\left(0,\; B_k - N_k \right)
\tag{25}
\end{equation}
Substituting:
\begin{equation}
B_{k+1} = \max\left(0,\; B_k - (1 - u_k)\frac{V_0}{1 + \gamma D_k} \right)
\tag{26}
\end{equation}
Thus, the evolution of the future backlog depends on the organization's base velocity $V_0$, on the decision regarding the priority assigned to debt remediation ($u_k$), and also on the performance degradation derived from the magnitude of the current technical debt.
Under these conditions, the total number of sprints required to complete the project is defined as:
\begin{equation}
K^{\ast} = \min \left\{ K_B \ge 0 \;:\; B_K = 0 \right\}
\tag{27}
\end{equation}
Equivalently, the completion condition can be expressed as:
\begin{equation}
\sum_{k=0}^{K^{\ast}-1} N_k \ge B_0
\tag{28}
\end{equation}
which means that what has been developed by the sprint $k$ under consideration exceeds the total backlog $B_0$; this condition can be written as:
\begin{equation}
\sum_{k=0}^{K^{\ast}-1} (1 - u_k)\frac{V_0}{1 + \gamma D_k} \ge B_0
\tag{29}
\end{equation}
The project cost discussed previously is therefore modified as:
\begin{equation}
\mathcal{C}^{\ast} = K^{\ast} \cdot C_{SP}
\tag{30}
\end{equation}
and given that $K^*>K$, it can be asserted that $C^*>C_0$, that is, in the absence of additional management measures, the existence of technical debt increases the total cost of the project, as expected given its condition as waste. This impact can be measured by:
\begin{equation}
\Delta K = K^{\ast} - K_B
\tag{31}
\end{equation}
\begin{equation}
\Delta \mathcal{C} = (K^{\ast} - K_B)\cdot p \cdot CPE \cdot t_{SP}
\tag{32}
\end{equation}
As stated, the evolution can be calculated at each sprint $k$, but for a preliminary analysis an approximation based on average values is used:
\begin{equation}
\bar{N} \approx (1 - \bar{u}) \frac{V_0}{1 + \gamma \bar{D}}
\tag{33}
\end{equation}
\begin{equation}
K^{\ast} \approx \left\lceil \frac{B_0 (1 + \gamma \bar{D})}{(1 - \bar{u}) V_0} \right\rceil
\tag{34}
\end{equation}
and the cost:
\begin{equation}
\mathcal{C}^{\ast} \approx 
\left\lceil \frac{B_0 (1 + \gamma \bar{D})}{(1 - \bar{u}) V_0} \right\rceil
\cdot p \cdot CPE \cdot t_{SP}
\tag{35}
\end{equation}
At the same time, the average variation in debt can be approximated as:
\begin{equation}
\Delta D \approx \left[\alpha(1 - \bar{u}) - (1 - \beta)\bar{u} \right] \bar{V}
\tag{36}
\end{equation}
The equilibrium condition ($\Delta D = 0$) leads to:
\begin{equation}
u_{\text{min}} = \frac{\alpha}{\alpha + 1 - \beta}
\tag{37}
\end{equation}
If $\bar{u} < u_{\text{min}}$, debt grows; if $\bar{u} > u_{\text{min}}$, debt tends to decrease. Project management should therefore use management actions to maintain debt sustainability by seeking to maintain the equilibrium condition.

The budget-viability condition is also modified to reflect the increase, so that:
\begin{equation}
K^{\ast} \cdot p \cdot CPE \cdot t_{SP} \le B
\tag{38}
\end{equation}
That is, technical debt impacts horizon and budget through three coupled mechanisms:

\begin{itemize}
\item generation of additional debt: $\alpha N_k$;
\item imperfect remediation of previous debt: $(1 - \beta) R_k$;
\item productivity degradation: $V_k = \frac{V_0}{1 + \gamma D_k}$.
\end{itemize}

The combined effect produces an increase in the number of required sprints and, consequently, a linear increase in the total project cost, or alternatively a reduction of project scope (reducing $K^*$) in order to satisfy the budget constraint.

\subsubsection{Use of Phase Containment of Errors (PCE)}

The differentiation between escaped defects on new functionality ($\alpha$) and imperfect remediation ($\beta$) makes it possible to model differential criteria for both processes. However, it is possible to approximate the behavior using a single organizational metric that captures, on average, defects escaping from one process phase to subsequent ones. This metric is usually called \textit{Phase Containment of Errors} (PCE) in the literature and is defined as:

\begin{equation}
PCE=\frac{defects_{in\ phase}}{defects_{in\ phase}+defects_{escaped}}
\tag{39}
\end{equation}
This metric is relatively simple to measure and usually converges to a stable metric for a given organizational development context, since it mainly depends on the methodology used. Then:
\begin{equation}
\alpha \approx (1-PCE)
\tag{40}
\end{equation}
and
\begin{equation}
1-\beta \approx PCE
\tag{41}
\end{equation}
Consequently, the structural sustainability threshold can be reduced to:
\begin{equation}
u_{min}=\frac{\alpha}{\alpha+1-\beta}=1-PCE
\tag{42}
\end{equation}

\section{Construction of the Debt-Management Policy}

From the discussion in the preceding sections, a model emerges to quantitatively characterize the qualitative notion that the appearance of involuntary debt deteriorates the initial scope estimate. It therefore allows preliminary planning to introduce organizational metrics derived from historical performance, such as $\alpha$, $\beta$, or $\gamma$, in order to plan the required budget or the obtainable scope more realistically.
However, the existence of technical debt creates the need to decide whether and how to solve it; that is, how to manage the remediation fraction $u_k$ in each sprint $k$. The model is deepened to obtain criteria that make it possible to interpret different strategies for managing technical debt.

\subsection{Construction of the naive policy}

The usual guideline dictated by good software-engineering practice is to address defects as soon as possible after their injection, because postponing their remediation can increase the future effort required and generate accumulated technical interest \cite{mcconnell,kruchten2012technical,kamei2016interest,ampatzoglou2015interest}.
Therefore, as a first approximation, if the idea is adopted that debt is the main problem and must be reduced as quickly as possible, the natural conclusion is to first allocate all available capacity to remediation until the debt accessible in the sprint is exhausted, and only then apply any remaining capacity to new functionality.
This policy, which we call \textit{naive}, prioritizes technical-debt remediation to the maximum possible extent in each sprint:
\begin{equation}
u_k =
\begin{cases}
1, & \text{if } D_k \ge V_k, \\[6pt]
\dfrac{D_k}{V_k}, & \text{if } 0 < D_k < V_k
\end{cases}
\tag{43}
\end{equation}

This induces two regimes:

\paragraph{Regime 1: dominant debt ($D_k \ge V_k$)}
\begin{equation}
N_k = 0, \qquad R_k = V_k
\tag{44}
\end{equation}
\begin{equation}
B_{k+1} = B_k
\tag{45}
\end{equation}
\paragraph{Regime 2: bounded debt ($D_k < V_k$)}
\begin{equation}
N_k = V_k - D_k, \qquad R_k = D_k
\tag{46}
\end{equation}
\begin{equation}
B_{k+1} = B_k - (V_k - D_k)
\tag{47}
\end{equation}

Therefore, backlog can be reduced even in the presence of technical debt, provided that:
\begin{equation}
V_k - D_k > 0
\tag{48}
\end{equation}
Under the naive policy, three structural features are observed. First, debt decreases monotonically while $D_k>0$, because all sprint capacity is assigned to it. Second, effective velocity gradually improves, as stated before a reduction in $D_k$ increases $V_k$. Third, the pure backlog does not change until the remaining debt is less than the attainable velocity. This implies a central economic consequence: although the naive policy improves the technical state of the system, it significantly postpones the capture of value associated with new functionalities.

It is therefore useful to study a set of heuristic policies that define different capacity-allocation strategies between new functionality and technical-debt remediation as possible alternatives for mitigating the value-impact consequences shown by the naive policy.
Let, as in the general model:
\begin{equation}
u_k \in [0,1]
\tag{49}
\end{equation}
be the fraction of sprint $k$ capacity dedicated to remediation; different policies are explored regarding how to define $u_k$ in a given project context for sprint $k$.

\subsubsection{\emph{Feature-first} policy}

This policy systematically prioritizes the development of new functionality, relegating debt remediation to a secondary or nonexistent role.
\begin{equation}
u_k = 0
\tag{50}
\end{equation}
Under this policy, the dynamics are:
\begin{equation}
N_k = V_k,\quad R_k = 0
\tag{51}
\end{equation}
which implies potential debt growth:
\begin{equation}
D_{k+1} = D_k + \alpha V_k
\tag{52}
\end{equation}
This policy maximizes the flow of new functionalities in the short term, but it may rapidly deteriorate future productivity through the deterioration of velocity ($V_k$) as a result of increasing debt, up to and including the condition of \textit{technical bankruptcy}.

\subsubsection{Fixed allocation}

In this policy, the fraction of capacity dedicated to remediation is constant over time and is chosen arbitrarily as a management decision.
\begin{equation}
u_k = \bar{u}, \quad \bar{u} \in (0,1)
\tag{53}
\end{equation}
Therefore:
\begin{equation}
N_k = (1-\bar{u})V_k,\quad R_k = \bar{u}V_k
\tag{54}
\end{equation}
This strategy is simple to implement, but it does not respond to changes in the system state $(B_k,D_k)$ nor does it explicitly capture value-optimization conditions.

\subsubsection{Threshold policy}

This policy establishes a critical debt level $D^{\ast}$ above which intensive remediation is activated:

\begin{equation}
u_k =
\begin{cases}
0 & \text{if } D_k < D^{\ast} \\
1 & \text{if } D_k \ge D^{\ast}
\end{cases}
\tag{55}
\end{equation}

This strategy introduces discontinuous behavior and may generate cycles of debt accumulation and discharge.

\subsubsection{Debt-proportional policy}

In this policy, the remediation fraction is proportional to the level of debt:
\begin{equation}
u_k = \min\left(1,\; \eta D_k \right)
\tag{56}
\end{equation}
where:
\begin{equation}
\eta > 0
\tag{57}
\end{equation}
is a sensitivity parameter.
This policy introduces a smooth response to the level of debt and tends to stabilize the system around a moderate debt level, seeking to avoid significant deterioration in project performance as a result of the parameter $\gamma$.

\subsubsection{Target-velocity policy}

In this case, the aim is to keep effective velocity above a threshold $V^{\ast}$.
Given that:
\begin{equation}
V_k = \frac{V_0}{1+\gamma D_k}
\tag{58}
\end{equation}
the policy can be formulated as:
\begin{equation}
u_k =
\begin{cases}
0 & \text{if } V_k \ge V^{\ast} \\
1 & \text{if } V_k < V^{\ast}
\end{cases}
\tag{59}
\end{equation}
or in continuous form:

\begin{equation}
u_k = \min\left(1,\; \xi (V^{\ast} - V_k)^{+} \right)
\tag{60}
\end{equation}
where $\xi>0$ is a sensitivity-adjustment parameter.

\subsubsection{Cost-based policy}

This policy approximates the optimal criterion by comparing marginal costs without explicitly solving the dynamic problem.

Let:
\begin{equation}
C_B(k) = \mu_B(B_k)
\tag{61}
\end{equation}

be the marginal cost of backlog, and

\begin{equation}
C_D(k) = MC_D(k)
\tag{62}
\end{equation}
the marginal cost of debt.

The policy is defined as:

\begin{equation}
u_k =
\begin{cases}
0 & \text{if } C_B(k) > (\alpha+1-\beta)C_D(k) \\
1 & \text{if } C_B(k) \le (\alpha+1-\beta)C_D(k)
\end{cases}
\tag{63}
\end{equation}
or in continuous form:
\begin{equation}
u_k =
\frac{(\alpha+1-\beta)C_D(k)}{C_B(k)+(\alpha+1-\beta)C_D(k)}
\tag{64}
\end{equation}
This policy constitutes a direct approximation to marginal equilibrium and anticipates the form of the optimal policy derived later.

\subsubsection{Comparative discussion}

The previous policies can be interpreted as progressively more sophisticated approximations to the capacity-allocation problem:

\begin{itemize}
\item \emph{Feature-first} and \textit{naive} represent opposite extremes.
\item Fixed allocation ignores the state of the system.
\item The threshold policy introduces reactive control.
\item The proportional policy smooths the response.
\item The target-velocity policy indirectly controls productivity.
\item The cost-based policy approximates the optimal economic criterion.
\end{itemize}

This set of policies makes it possible to understand the conceptual transition from heuristic rules toward the rigorous formulation of the optimal-control problem developed in the following sections.
Therefore, the correct decision cannot be formulated as: ``eliminate debt at the maximum rate.'' It must be formulated as: ``determine how much of sprint capacity should be assigned to debt and how much to new functions.'' This leads to the general case.

\section{General Case}

To address a resource-allocation policy ($u_k$) that defines what proportion of resources can be associated with solving backlog and what proportion with debt remediation, we build from the basic economic objects of the sprint.

The formulation must respect three fundamental conditions:

\begin{enumerate}[label=\arabic*.]
\item The variables $B_k$, $D_k$, $N_k$, $R_k$, and $V_k$ are measured in effort units, for example story points, and not in number of stories.
\item The policy $u_k$ decides the partition of the effective sprint velocity, not the direct selection of individual items.
\item If no technical debt is available, that is, if $D_k=0$, the policy must necessarily satisfy $u_k^\star=0$.
\end{enumerate}

The economic problem of the sprint is to compare:

\begin{enumerate}[label=\arabic*.]
\item the value of applying $N_k$ story points to the backlog;
\item the value of applying $R_k$ story points to debt;
\item the future productivity gain derived from reducing debt.
\end{enumerate}

The optimal decision is not made item by item, but over stocks of effort units (story points). Therefore, it is not enough to compare ``\textit{the next story}'' in the backlog with ``\textit{the next story}'' in debt. The model must compare aggregate marginal values per unit of effort.

Before formally defining the economic value of the pure backlog, it is necessary to make explicit the assumption under which the backlog ranking can be used as a proxy for marginal value.

\subsection{Reasonably well-prioritized backlog assumption}

The economic formulation of backlog adopted in this work assumes that the backlog is prioritized in a way reasonably consistent with the expected economic value of its items. In particular, the representation through a decreasing marginal-value function should not be interpreted as a claim that the prioritization observed in an agile project is perfect, stable, or fully objective, but rather as an aggregate approximation that makes it possible to transform the remaining backlog into a tractable economic stock.

This assumption is relevant because the model associates progress over the pure backlog $B_k$ with progressive value capture, and compares that marginal value with the economic value of remediating functional debt $D_k$. Therefore, the allocation policy between new functionalities and remediation depends on the relative order of the backlog preserving, at least approximately, useful information about value. In practical terms, it is assumed that the process of backlog refinement, prioritization, and review has produced an ordering sufficiently good for items located in higher-priority positions to have, on average, greater expected value than items located in later positions.

In real Scrum projects, this assumption may fail for several reasons. Prioritization may be noisy, affected by political pressures, respond to local urgencies, reflect preferences of dominant actors, or change endogenously as technical progress reveals initially unobserved restrictions, dependencies, or risks. In such cases, the backlog ranking ceases to be a clean signal of marginal value and the value function used by the model must be interpreted with caution.

If prioritization is noisy but unbiased, the model may still be useful as a first-order approximation, although the estimated marginal quantities will be subject to greater variance. If prioritization is systematically biased, for example because low-value items are advanced for political reasons or technically critical items are postponed, then the resulting policy may induce suboptimal decisions: it may overestimate the value of developing new functionalities or underestimate the convenience of remediating technical debt. Finally, if prioritization is endogenous to the technical state of the system itself, the ranking should be treated as an additional dynamic variable and not as a fixed structure.

Consequently, the model should be read under the explicit hypothesis of a reasonably well-prioritized backlog. When this condition is not met, a natural extension is to introduce uncertainty in the ranking, correction factors on marginal value, or a reprioritization mechanism dependent on the technical state $(B_k,D_k,V_k)$. These extensions do not invalidate the basic structure of the model, but shift its interpretation from a deterministic decision rule toward an analytical tool under organizational and economic uncertainty using stochastic modelling with risk analysis tools.

\subsection{Backlog value structure}

It is assumed that functionalities are prioritized and that their value follows a Pareto/Zipf law:
\begin{equation}
v(i)=A i^{-s}
\tag{65}
\end{equation}
where:

\begin{itemize}
\item $i$ is the priority rank;
\item $A>0$ is the value of the most valuable functionality;
\item $s>0$ measures value concentration.
\end{itemize}

But $B_k$ is not the number of stories. It is remaining effort in story points. Therefore, we introduce:
\begin{equation}
\bar s_B>0
\tag{66}
\end{equation}
as the average size, in story points, of a backlog story.

If the remaining backlog at the beginning of the sprint is $B_k$, the approximate number of equivalent backlog stories is:
\begin{equation}
\frac{B_k}{\bar s_B}
\tag{67}
\end{equation}
If $N_k$ story points are applied to backlog, this is approximately equivalent to executing:
\begin{equation}
\frac{N_k}{\bar s_B}
\tag{68}
\end{equation}
equivalent stories from the head of the prioritized backlog.

The value captured by executing $N_k$ story points of backlog is approximated as a continuous integral:
\begin{equation}
\mathcal V_B(N_k;B_k)
=
\frac{A}{\bar s_B}
\int_0^{N_k}
\left(
M-\frac{B_k}{\bar s_B}+1+\frac{x}{\bar s_B}
\right)^{-s} dx
\tag{69}
\end{equation}
This formula says the following:
\begin{itemize}
\item $M$ represents the initial number of equivalent stories in the portfolio;
\item $M-\frac{B_k}{\bar s_B}+1$ approximates the rank of the first pending backlog story;
\item $\frac{x}{\bar s_B}$ advances in the ranking as story points are consumed;
\item the factor $\frac{A}{\bar s_B}$ converts value per equivalent story into value per story point.
\end{itemize}

The marginal value of backlog is the additional value obtained by dedicating one more story point to new functionality:
\begin{equation}
\mu_B(N_k;B_k)
=
\frac{\partial \mathcal V_B}{\partial N}
\tag{70}
\end{equation}
Deriving (69):
\begin{equation}
\mu_B(N_k;B_k)
=
\frac{A}{\bar s_B}
\left(
M-\frac{B_k}{\bar s_B}+1+\frac{N_k}{\bar s_B}
\right)^{-s}
\tag{71}
\end{equation}

This term is not the value of ``one story,'' but the marginal value per backlog story point, evaluated after assigning $N_k$ story points to new functionality.
For an initial sprint approximation, it is evaluated at $N_k=0$:
\begin{equation}
\mu_B^{(0)}(B_k)
=
\frac{A}{\bar s_B}
\left(
M-\frac{B_k}{\bar s_B}+1
\right)^{-s}
\tag{72}
\end{equation}
This is the \textit{marginal value of backlog} at the beginning of sprint $k$.

\subsection{Debt value structure}

Technical debt represents previously prioritized work that did not successfully complete the development cycle. Since it comes from functionalities that were addressed earlier, it is located in a higher-priority zone than the remaining backlog.
But there is an essential difference: functionality in debt may have already captured part of its value. Therefore, we introduce:
\begin{equation}
\theta_k\in[0,1]
\tag{73}
\end{equation}
where $\theta_k$ is the average fraction of value already captured by the items in debt.
Then:
\begin{equation}
1-\theta_k
\tag{74}
\end{equation}
is the fraction of residual value recoverable when remediating debt.
We also introduce:
\begin{equation}
\bar s_D>0
\tag{75}
\end{equation}
as the average size, in story points, of debt units.

If $D_k$ are story points of debt, the approximate number of equivalent stories in debt is:
\begin{equation}
\frac{D_k}{\bar s_D}
\tag{76}
\end{equation}
If $R_k$ story points are applied to remediation, the residual value recovered is approximated by:
\begin{equation}
\mathcal V_D(R_k;B_k,D_k)
=
(1-\theta_k)
\frac{A}{\bar s_D}
\int_0^{R_k}
\left(
M-\frac{B_k}{\bar s_B}
-\frac{D_k}{\bar s_D}
+1+\frac{x}{\bar s_D}
\right)^{-s} dx
\tag{77}
\end{equation}
The position:
\begin{equation}
M-\frac{B_k}{\bar s_B}
-\frac{D_k}{\bar s_D}
+1
\tag{78}
\end{equation}
represents the approximate beginning of the debt band in the value ranking.
The marginal value of debt is the additional recoverable value from dedicating one more story point to remediation:
\begin{equation}
\mu_D(R_k;B_k,D_k)
=
\frac{\partial \mathcal V_D}{\partial R}
\tag{79}
\end{equation}
Deriving (77):
\begin{equation}
\mu_D(R_k;B_k,D_k)
=
(1-\theta_k)
\frac{A}{\bar s_D}
\left(
M-\frac{B_k}{\bar s_B}
-\frac{D_k}{\bar s_D}
+1+\frac{R_k}{\bar s_D}
\right)^{-s}
\tag{80}
\end{equation}
For the initial sprint approximation, we evaluate at $R_k=0$:
\begin{equation}
\mu_D^{(0)}(B_k,D_k)
=
(1-\theta_k)
\frac{A}{\bar s_D}
\left(
M-\frac{B_k}{\bar s_B}
-\frac{D_k}{\bar s_D}
+1
\right)^{-s}
\tag{81}
\end{equation}

This is the \textit{marginal value of debt} at the beginning of sprint $k$.

\subsection{Gain from productivity improvement}

Debt not only has residual value; it also reduces velocity according to equation [4], and therefore remediating it may increase future velocity.
The future velocity attributable to that remediation is approximated as:
\begin{equation}
V_{k+1}^{rem}(R_k)
=
\frac{V_0}{1+\gamma\left(D_k-(1-\beta)R_k\right)}
\tag{82}
\end{equation}
The velocity improvement generated by remediation is:
\begin{equation}
\Delta V_{k+1}(R_k)
=
\frac{V_0}{1+\gamma\left(D_k-(1-\beta)R_k\right)}
-
\frac{V_0}{1+\gamma D_k}
\tag{83}
\end{equation}

\subsubsection{Meaning of $\lambda_k$}

Future velocity improvement has economic value because it makes it possible to execute more backlog in the future. To monetize this improvement, we introduce:
\begin{equation}
\lambda_k>0
\tag{84}
\end{equation}
as the marginal economic value of one additional story point of future capacity.
Formally, it can be understood as a shadow price:
\begin{equation}
\lambda_k=\frac{\partial J_k}{\partial V_{k+1}}
\tag{85}
\end{equation}
where $J_k$ is the future economic value of the project from sprint $k$ onward.
With this, the economic gain from productivity improvement is:
\begin{equation}
\mathcal G_D(R_k;D_k)
=
\lambda_k \Delta V_{k+1}(R_k)
\tag{86}
\end{equation}
Substituting (83):
\begin{equation}
\mathcal G_D(R_k;D_k)
=
\lambda_k
\left[
\frac{V_0}{1+\gamma\left(D_k-(1-\beta)R_k\right)}
-
\frac{V_0}{1+\gamma D_k}
\right]
\tag{87}
\end{equation}
The marginal productivity gain from one additional story point of remediation is:
\begin{equation}
g_D(R_k;D_k)
=
\frac{\partial \mathcal G_D}{\partial R}(R_k;D_k)
\tag{88}
\end{equation}
Deriving (87):
\begin{equation}
g_D(R_k;D_k)
=
\lambda_k
\frac{
V_0\gamma(1-\beta)
}{
\left[
1+\gamma\left(D_k-(1-\beta)R_k\right)
\right]^2
}
\tag{89}
\end{equation}
The initial sprint approximation is obtained by evaluating at $R_k=0$:
\begin{equation}
g_D^{(0)}(D_k)
=
\lambda_k
\frac{
V_0\gamma(1-\beta)
}{
(1+\gamma D_k)^2
}
\tag{90}
\end{equation}
Debt therefore has two marginal benefits:

\begin{enumerate}[label=\arabic*.]
\item recovered residual value;
\item future productivity improvement.
\end{enumerate}

Therefore, the total marginal value of remediating debt at the beginning of sprint $k$ is:
\begin{equation}
m_D^{(0)}(B_k,D_k)
=
\mu_D^{(0)}(B_k,D_k)+g_D^{(0)}(D_k)
\tag{91}
\end{equation}
Substituting:
\begin{align}
\mu_D^{(0)}(B_k,D_k)
&=
(1-\theta_k)
\frac{A}{\bar s_D}
\left(
M-\frac{B_k}{\bar s_B}
-\frac{D_k}{\bar s_D}
+1
\right)^{-s}
\tag{92}\\
g_D^{(0)}(D_k)
&=
\lambda_k
\frac{
V_0\gamma(1-\beta)
}{
(1+\gamma D_k)^2
}
\tag{93}\\
m_D^{(0)}(B_k,D_k)
&=
\mu_D^{(0)}(B_k,D_k)+g_D^{(0)}(D_k)
\tag{94}
\end{align}

\subsection{Unrestricted economic policy}

The pure economic policy, before considering boundary restrictions of applicability, distributes capacity in proportion to marginal attractiveness at the beginning of each sprint $k$:
\begin{equation}
\widehat u_k
=
\frac{
m_D^{(0)}(B_k,D_k)
}{
\mu_B^{(0)}(B_k)+m_D^{(0)}(B_k,D_k)
}
\tag{95}
\end{equation}
Substituting (72) and (91):
\begin{align}
Z_k
&=
(1-\theta_k)
\dfrac{A}{\bar s_D}
\left(
M-\dfrac{B_k}{\bar s_B}
-\dfrac{D_k}{\bar s_D}
+1
\right)^{-s}
+
\lambda_k
\dfrac{
V_0\gamma(1-\beta)
}{
(1+\gamma D_k)^2
}
\tag{96}\\
Y_k
&=
\dfrac{A}{\bar s_B}
\left(
M-\dfrac{B_k}{\bar s_B}
+1
\right)^{-s}
\tag{97}\\
Q_k
&=
Y_k+Z_k
\tag{98}\\
\widehat u_k
&=
\frac{Z_k}{Q_k}
\tag{99}
\end{align}

With the assumptions and simplifications adopted so far, this formulation predicts $\widehat u_k$. However, in the case $D_k>0$ it may allocate capacity to debt even when physically nothing can be remediated because debt does not exist. In general, even when debt exists, no more debt can be remediated than is available.

The physical constraint is:
\begin{equation}
0\le R_k\le D_k
\tag{100}
\end{equation}
Since:
\begin{equation}
R_k=u_kV_k
\tag{101}
\end{equation}
we obtain:
\begin{equation}
0\le u_kV_k\le D_k
\tag{102}
\end{equation}
and therefore:
\begin{equation}
0\le u_k\le \frac{D_k}{V_k}
\tag{103}
\end{equation}
Since, in addition, $u_k\le1$, the feasible set is:
\begin{equation}
0\le u_k\le u_k^{\max}
\tag{104}
\end{equation}
where:
\begin{equation}
u_k^{\max}
=
\min\left(1,\frac{D_k}{V_k}\right)
\tag{105}
\end{equation}
Substituting $V_k$:
\begin{equation}
u_k^{\max}
=
\min\left(
1,
\frac{D_k(1+\gamma D_k)}{V_0}
\right)
\tag{106}
\end{equation}

The final policy does not eliminate the pure economic formula, but projects it onto the feasible interval, which means:
\begin{equation}
u_k^\star=
\begin{cases}
0, & D_k=0 \\[6pt]
\min\left(\widehat u_k,u_k^{\max}\right), & D_k>0
\end{cases}
\tag{107}
\end{equation}
Equivalently:
\begin{equation}
u_k^\star=
\begin{cases}
0, & D_k=0 \\[6pt]
\min\left[
\widehat u_k,
\min\left(
1,
\dfrac{D_k(1+\gamma D_k)}{V_0}
\right)
\right], & D_k>0
\end{cases}
\tag{108}
\end{equation}
This completes the formulation of the general case.

\subsection{Simplified case}

The general case can be simplified under certain conditions, particularly when it can be assumed that:
\begin{equation}
\gamma \approx 0
\tag{109}
\end{equation}
there is no significant degradation of velocity with accumulated debt.
\begin{equation}
\theta_k \approx 0
\tag{110}
\end{equation}
The value realized by functions in the debt stock $D_k$ is relatively low.
\begin{equation}
\lambda_k \approx 0
\tag{111}
\end{equation}
There are no significant differences in the average effort between backlog and debt items.
\begin{equation}
\bar s_D=\bar s_B=\bar s
\tag{112}
\end{equation}
and also:
\begin{equation}
\alpha=1-PCE
\tag{113}
\end{equation}
\begin{equation}
1-\beta=PCE
\tag{114}
\end{equation}
When the simplification is applied:
\begin{equation}
V_k=V_0
\tag{115}
\end{equation}
and therefore the gain from future productivity disappears:
\begin{equation}
g_D^{(0)}=0
\tag{116}
\end{equation}
Since:
\begin{equation}
\theta_k=0
\tag{117}
\end{equation}
all debt value is considered residual and recoverable:
\begin{equation}
1-\theta_k=1
\tag{118}
\end{equation}
And since:
\begin{equation}
\bar s_D=\bar s_B=\bar s
\tag{119}
\end{equation}
the average sizes of backlog and debt are equal.

The initial marginal value of the backlog becomes:
\begin{equation}
\mu_B^{(0)}
=
\frac{A}{\bar s}
\left(
M-\frac{B_k}{\bar s}+1
\right)^{-s}
\tag{120}
\end{equation}
The marginal value of debt at the beginning of sprint $k$ becomes:
\begin{equation}
\mu_D^{(0)}
=
\frac{A}{\bar s}
\left(
M-\frac{B_k+D_k}{\bar s}+1
\right)^{-s}
\tag{121}
\end{equation}
Since $g_D^{(0)}=0$, the equilibrium condition reduces to:
\begin{equation}
m_D^{(0)}=\mu_D^{(0)}
\tag{122}
\end{equation}
Therefore, the unrestricted policy is:
\begin{equation}
\widehat u_k
=
\frac{
\mu_D^{(0)}
}{
\mu_B^{(0)}+\mu_D^{(0)}
}
\tag{123}
\end{equation}
Substituting:
\begin{equation}
\widehat u_k
=
\frac{
\dfrac{A}{\bar s}
\left(
M-\dfrac{B_k+D_k}{\bar s}+1
\right)^{-s}
}{
\dfrac{A}{\bar s}
\left(
M-\dfrac{B_k}{\bar s}+1
\right)^{-s}
+
\dfrac{A}{\bar s}
\left(
M-\dfrac{B_k+D_k}{\bar s}+1
\right)^{-s}
}
\tag{124}
\end{equation}

Canceling $\frac{A}{\bar s}$:

\begin{equation}
\widehat u_k
=
\frac{
\left(
M-\dfrac{B_k+D_k}{\bar s}+1
\right)^{-s}
}{
\left(
M-\dfrac{B_k}{\bar s}+1
\right)^{-s}
+
\left(
M-\dfrac{B_k+D_k}{\bar s}+1
\right)^{-s}
}
\tag{125}
\end{equation}
This is the simplified deterministic formula for $D_k>0$, before applying the boundary. When doing so, the feasibility bound (existence of debt) is:
\begin{equation}
u_k^{\max}
=
\min\left(1,\frac{D_k}{V_0}\right)
\tag{126}
\end{equation}
Therefore, the final simplified policy is:
\begin{equation}
u_k^\star=
\begin{cases}
0, & D_k=0 \\[6pt]
\min\left(\widehat u_k,\min\left(1,\dfrac{D_k}{V_0}\right)\right), & D_k>0
\end{cases}
\tag{127}
\end{equation}

where $\widehat u_k$ is given as before.

The minimum debt-sustainability condition remains:
\begin{equation}
u_k^\star\ge1-PCE
\tag{128}
\end{equation}

\section{Discrete Extension with Inhomogeneous Sizes and Item Indivisibility}

\setcounter{equation}{131}

The continuous model previously developed assumes that effort can be fractionally allocated between technical debt and backlog. However, in practice items are indivisible and have heterogeneous sizes. This introduces a discrepancy between the optimal continuous solution and effective implementation in a sprint.

We start from the formulation of value affected by technical debt:

\begin{equation}
V_k=\frac{V_0}{1+\gamma D_k}
\tag{129}
\end{equation}

The continuous policy determines an optimal effort assigned to debt:

\begin{equation}
\widehat R_k = u_k^\star V_k
\tag{130}
\end{equation}

However, technical debt is not divisible. There exists a discrete set of items:

\begin{equation}
\mathcal D_k=\{d_{k,1},d_{k,2},\dots,d_{k,n_D}\}
\tag{131}
\end{equation}

with sizes:

\begin{equation}
q^D_{k,i}>0
\tag{132}
\end{equation}

marginal values ordered according to a Pareto/Zipf-like distribution:

\begin{equation}
v^D_{k,1}\ge v^D_{k,2}\ge \dots \ge v^D_{k,n_D}
\tag{133}
\end{equation}

Analogously, for the backlog:

\begin{equation}
\mathcal B_k=\{b_{k,1},b_{k,2},\dots,b_{k,n_B}\}
\tag{134}
\end{equation}

\begin{equation}
q^B_{k,j}>0
\tag{135}
\end{equation}

\begin{equation}
v^B_{k,1}\ge v^B_{k,2}\ge \dots \ge v^B_{k,n_B}
\tag{136}
\end{equation}

Effort allocation becomes discrete:

\begin{equation}
R_k = \sum_{i=1}^{n_D} x^D_{k,i} q^D_{k,i}
\tag{137}
\end{equation}

\begin{equation}
N_k = \sum_{j=1}^{n_B} x^B_{k,j} q^B_{k,j}
\tag{138}
\end{equation}

where:

\begin{equation}
x^D_{k,i},x^B_{k,j}\in\{0,1\}
\tag{139}
\end{equation}

and the capacity constraint is:

\begin{equation}
\sum_i x^D_{k,i} q^D_{k,i}
+
\sum_j x^B_{k,j} q^B_{k,j}
\le V_k
\tag{140}
\end{equation}

The discrete problem is formulated as:

\begin{equation}
\max_{x^D,x^B}
\left[
\sum_i x^D_{k,i} w^D_{k,i}
+
\sum_j x^B_{k,j} w^B_{k,j}
\right]
\tag{141}
\end{equation}

where the economic values are:

\begin{equation}
w^D_{k,i}
=
\text{recovered residual value}
+
\text{future productivity gain}
\tag{142}
\end{equation}

\begin{equation}
w^B_{k,j}
=
\text{captured new functional value}
\tag{143}
\end{equation}

Under the Pareto hypothesis:

\begin{equation}
w^D_{k,i} \ge w^B_{k,j}
\tag{144}
\end{equation}

\subsection{Policy 1: Rounding toward debt (ceiling)}

\begin{equation}
m_D^\uparrow
=
\min\left\{
m:
\sum_{i=1}^{m} q^D_{k,i}
\ge
\widehat R_k
\right\}
\tag{145}
\end{equation}

\begin{equation}
R_k^{disc}
=
\sum_{i=1}^{m_D^\uparrow} q^D_{k,i}
\tag{146}
\end{equation}

\subsection{Policy 2: Floor on debt and remainder to backlog}

\begin{equation}
m_D^\downarrow
=
\max\left\{
m:
\sum_{i=1}^{m} q^D_{k,i}
\le
\widehat R_k
\right\}
\tag{147}
\end{equation}

\begin{equation}
R_k^{disc}
=
\sum_{i=1}^{m_D^\downarrow} q^D_{k,i}
\tag{148}
\end{equation}

\begin{equation}
L_k = V_k - R_k^{disc}
\tag{149}
\end{equation}

\begin{equation}
m_B^\downarrow
=
\max\left\{
m:
\sum_{j=1}^{m} q^B_{k,j}
\le
L_k
\right\}
\tag{150}
\end{equation}

\begin{equation}
N_k^{disc}
=
\sum_{j=1}^{m_B^\downarrow} q^B_{k,j}
\tag{151}
\end{equation}

\begin{equation}
W_k
=
V_k-R_k^{disc}-N_k^{disc}
\tag{152}
\end{equation}

\begin{equation}
0\le W_k < \min(q^D_{next},q^B_{next})
\tag{153}
\end{equation}

\subsection{Policy 3: Discrete knapsack-type optimum}

\begin{equation}
\max
\sum_i x^D_i w^D_i+\sum_j x^B_j w^B_j
\tag{154}
\end{equation}

\begin{equation}
\sum_i x^D_i q^D_i+\sum_j x^B_j q^B_j\le V_k
\tag{155}
\end{equation}

\begin{equation}
x^D_i,x^B_j\in\{0,1\}
\tag{156}
\end{equation}

The effective executed proportion is:

\begin{equation}
u_k^{disc}
=
\frac{
\sum_i x^D_{k,i}q^D_{k,i}
}{
V_k
}
\tag{157}
\end{equation}

The discretization error is:

\begin{equation}
\varepsilon_k
=
u_k^{disc}-u_k^\star
\tag{158}
\end{equation}

The wasted capacity is:

\begin{equation}
W_k
=
V_k
-
\sum_i x^D_{k,i} q^D_{k,i}
-
\sum_j x^B_{k,j} q^B_{k,j}
\tag{159}
\end{equation}

The packing efficiency is:

\begin{equation}
\eta_k
=
1-\frac{W_k}{V_k}
\tag{160}
\end{equation}

\subsection{Interpretation}

The continuous model produces an optimal marginal policy:

\begin{equation}
u_k^\star V_k \in \mathbb R
\tag{161}
\end{equation}

whereas the real implementation belongs to the discrete set:

\begin{equation}
R_k^{disc},N_k^{disc}
\in
\left\{
\sum q_i
\right\}
\tag{162}
\end{equation}

Therefore:

\begin{equation}
u_k^\star
\longrightarrow
(x^D_{k,i},x^B_{k,j})
\longrightarrow
u_k^{disc}
\tag{163}
\end{equation}

with:

\begin{equation}
u_k^{disc}
=
\frac{R_k^{disc}}{V_k}
\tag{164}
\end{equation}

This introduces a structural gap between the continuous optimum and discrete execution, characterized by:

\begin{itemize}
\item Discretization error $\varepsilon_k$;
\item Idle capacity $W_k$;
\item Packing efficiency $\eta_k$.
\end{itemize}

The refined model is thus defined at two levels:

\begin{itemize}
\item Continuous level: determines $u_k^\star$;
\item Discrete level: selects executable items.
\end{itemize}

The fundamental tension arises because:

\begin{equation}
u_k^\star V_k \notin \text{discrete feasible set}
\tag{165}
\end{equation}

\section{Model Validation}

To validate the model, runs are performed in different scenarios and the results are observed to determine whether they reasonably predict the empirical behavior observed in projects and the response to the hypotheses of systemic relationships.

To carry this out, a series of runs is performed under a simulated framework of values for the systemic variables and the sensitivity of the model's defined parameters.

\subsection{Systemic behavior}

The objective is to observe the behavior of $u_k$ for different backlog ($B_k$) and technical debt ($D_k$) situations using a parameter configuration assumed to be representative of real organizations based on available data and experience.

\subsubsection{Model normalization}

To facilitate execution, the derived model  is reformulated to use backlog and debt magnitudes normalized to the total backlog size and proportions between debt and backlog.

\begin{equation}
Z_k=(1-\theta_k)
\dfrac{A}{\bar s_D}
\left(
M-\dfrac{B_k}{\bar s_B}
-\dfrac{D_k}{\bar s_D}
+1
\right)^{-s}
+
\lambda_k
\dfrac{
V_0\gamma(1-\beta)
}{
(1+\gamma D_k)^2
}
\tag{166}
\end{equation}

\begin{equation}
Y_k=\dfrac{A}{\bar s_B}
\left(
M-\dfrac{B_k}{\bar s_B}
+1
\right)^{-s}
\tag{167}
\end{equation}

\begin{equation}
Q_k=Y_k+Z_k
\tag{168}
\end{equation}

\begin{equation}
\widehat u_k=\frac{Z_k}{Q_k}
\tag{169}
\end{equation}

\begin{equation}
\widehat u_k=\frac{
m_D^{(0)}(B_k,D_k)
}{
\mu_B^{(0)}(B_k)+m_D^{(0)}(B_k,D_k)
}
\tag{170}
\end{equation}

The sought normalization is therefore:
\begin{equation}
b_k=\frac{B_k/\bar s_B}{M}=\frac{B_k}{M\bar s_B}
\tag{171}
\end{equation}
\begin{equation}
d_k=\frac{D_k/\bar s_D}{M}=\frac{D_k}{M\bar s_D}
\tag{172}
\end{equation}
Therefore:
\begin{equation}
\frac{B_k}{\bar s_B}=M b_k
\tag{173}
\end{equation}
\begin{equation}
\frac{D_k}{\bar s_D}=M d_k
\tag{174}
\end{equation}

\begin{equation}
D_k=M\bar s_D d_k
\tag{175}
\end{equation}

Substituting in the marginal term associated with the pure backlog:
\begin{equation}
Y_k=\dfrac{A}{\bar s_B}
\left(
M-Mb_k+1
\right)^{-s}
\tag{176}
\end{equation}
\begin{equation}
Y_k^{(n)}=\dfrac{A}{\bar s_B}
\left(
M(1-b_k)+1
\right)^{-s}
\tag{177}
\end{equation}

Substituting now in the debt-remediation component:
\begin{equation}
(1-\theta_k)
\dfrac{A}{\bar s_D}
\left(
M-\dfrac{B_k}{\bar s_B}
-\dfrac{D_k}{\bar s_D}
+1
\right)^{-s}
\tag{178}
\end{equation}
we obtain:
\begin{equation}
(1-\theta_k)
\dfrac{A}{\bar s_D}
\left(
M-Mb_k-Md_k+1
\right)^{-s}
\tag{179}
\end{equation}

that is:
\begin{equation}
(1-\theta_k)
\dfrac{A}{\bar s_D}
\left(
M(1-b_k-d_k)+1
\right)^{-s}
\tag{180}
\end{equation}

The productivity-gain term requires more care, because it depends on $D_k$, not on $D_k/\bar s_D$. Thus:
\begin{equation}
\lambda_k
\dfrac{
V_0\gamma(1-\beta)
}{
(1+\gamma D_k)^2
}
\tag{181}
\end{equation}
becomes:
\begin{equation}
\lambda_k
\dfrac{
V_0\gamma(1-\beta)
}{
\left(1+\gamma M\bar s_D d_k\right)^2
}
\tag{182}
\end{equation}
Therefore:

\begin{equation}
Z_k^{(n)}=(1-\theta_k)\dfrac{A}{\bar s_D}\left(
M(1-b_k-d_k)+1
\right)^{-s}
+
\lambda_k
\dfrac{
V_0\gamma(1-\beta)
}{
\left(1+\gamma M\bar s_D d_k\right)^2
}
\tag{183}
\end{equation}
Then:

\begin{equation}
Q_k^{(n)}=Y_k^{(n)}+Z_k^{(n)}
\tag{184}
\end{equation}

and the continuous, complete, unrestricted, normalized policy is:

\begin{equation}
\widehat u_k^{(n)}=\frac{
Z_k^{(n)}
}{
Q_k^{(n)}
}
=
\frac{
Z_k^{(n)}
}{
Y_k^{(n)}+Z_k^{(n)}
}
\tag{185}
\end{equation}

Equivalent to equation 95:

\begin{equation}
\widehat u_k^{(n)}=
\frac{
m_D^{(0,n)}(b_k,d_k)
}{
\mu_B^{(0,n)}(b_k)+m_D^{(0,n)}(b_k,d_k)
}
\tag{186}
\end{equation}

where:

\begin{equation}
\mu_B^{(0,n)}(b_k)=
\dfrac{A}{\bar s_B}
\left(
M(1-b_k)+1
\right)^{-s}
\tag{187}
\end{equation}

\begin{equation}
\mu_D^{(0,n)}(b_k,d_k)=(1-\theta_k)
\dfrac{A}{\bar s_D}
\left(
M(1-b_k-d_k)+1
\right)^{-s}
\tag{188}
\end{equation}

\begin{equation}
g_D^{(0,n)}(d_k)=\lambda_k
\dfrac{
V_0\gamma(1-\beta)
}{
\left(1+\gamma M\bar s_D d_k\right)^2
}
\tag{189}
\end{equation}

\begin{equation}
m_D^{(0,n)}(b_k,d_k)=\mu_D^{(0,n)}(b_k,d_k)
+
g_D^{(0,n)}(d_k)
\tag{190}
\end{equation}
Strictly speaking, normalization does not eliminate $M$; it simply makes it easier to define scenarios that operate on proportions rather than concrete values.
\begin{equation}
\frac{B_k}{\bar s_B}
\rightarrow
M b_k
\tag{191}
\end{equation}
\begin{equation}
\frac{D_k}{\bar s_D}
\rightarrow
M d_k
\tag{192}
\end{equation}
but the economic productivity term depends on debt in original units:
\begin{equation}
D_k=M\bar s_D d_k
\tag{193}
\end{equation}
which is why the following appears:
\begin{equation}
1+\gamma M\bar s_D d_k
\tag{194}
\end{equation}
and not only:
\begin{equation}
1+\gamma d_k
\tag{195}
\end{equation}

unless an additional normalized parameter is defined:
\begin{equation}
\gamma_D^{(n)}=\gamma M\bar s_D
\tag{196}
\end{equation}
in which case:

\begin{equation}
g_D^{(0,n)}(d_k)=\lambda_k
\dfrac{
V_0\gamma(1-\beta)
}{
\left(1+\gamma_D^{(n)}d_k\right)^2
}
\tag{197}
\end{equation}

\subsubsection{Systemic behavior}

The behavior is explored for different proportions $B_k/M$, representing the remaining backlog at the time of analysis, and configurations of the ratio $D_k/B_k$ (the proportion of technical debt over the remaining backlog).

\begin{figure}[H]
    \centering
    \includegraphics[width=0.95\linewidth]{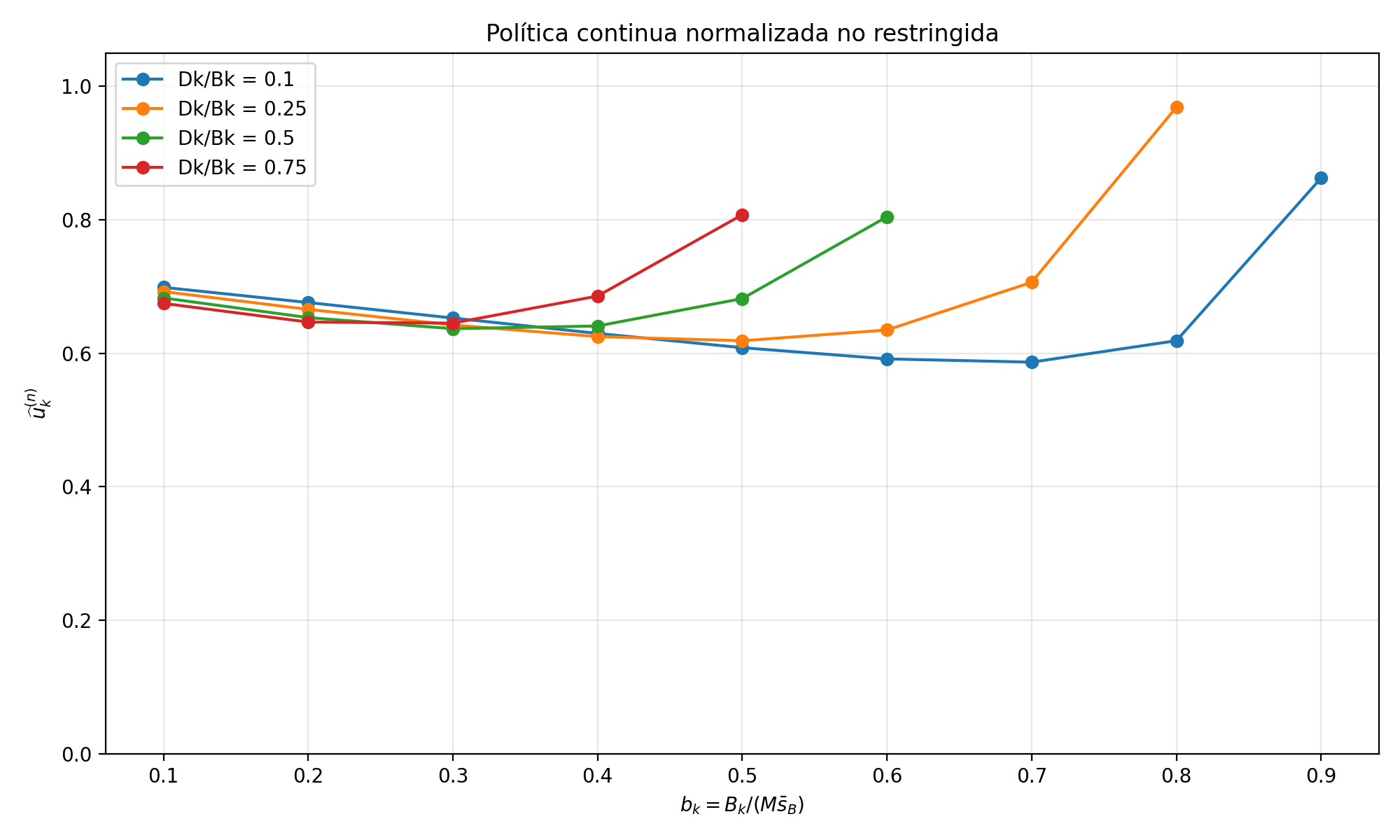}
    \caption{Normalized model as a function of $B_k/M$ and $D_k/B_k$}
    \label{fig:model_01}
\end{figure}

The model predicts that there will be some preference for resolving technical debt before progressing with backlog ($B_k/M\le0.3$).

\subsubsection{Sensitivity to the ratio $D_k/B_k$}

The model is used to evaluate the sensitivity to the ratio $D_k/B_k$.

\begin{figure}[H]
    \centering
    \includegraphics[width=0.95\linewidth]{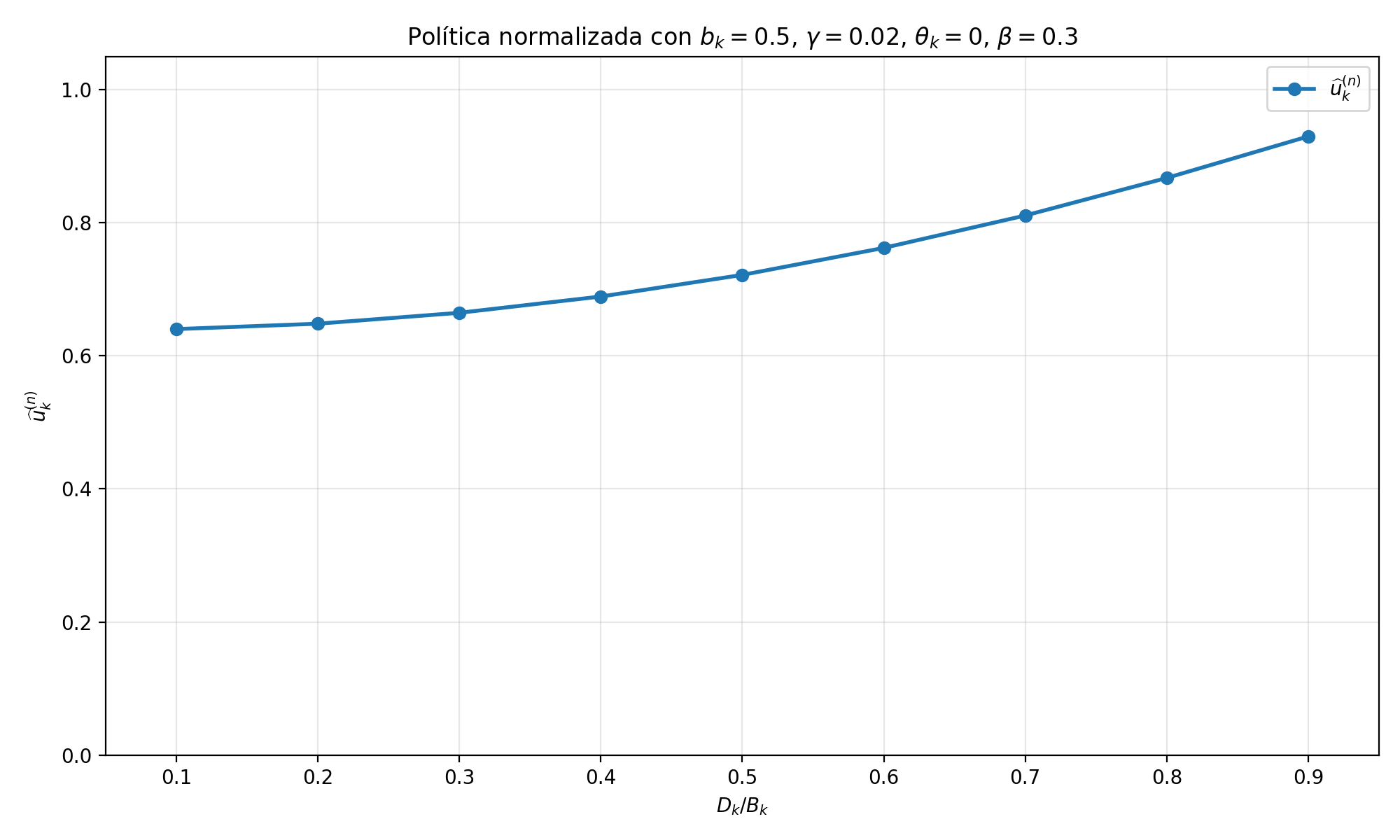}
    \caption{Sensitivity to the ratio $D_k/B_k$}
    \label{fig:model_02}
\end{figure}

\subsubsection{Sensitivity to variations in $\gamma$}

The model is used to evaluate the sensitivity to the $\gamma$ factor.

\begin{figure}[H]
    \centering
    \includegraphics[width=0.95\linewidth]{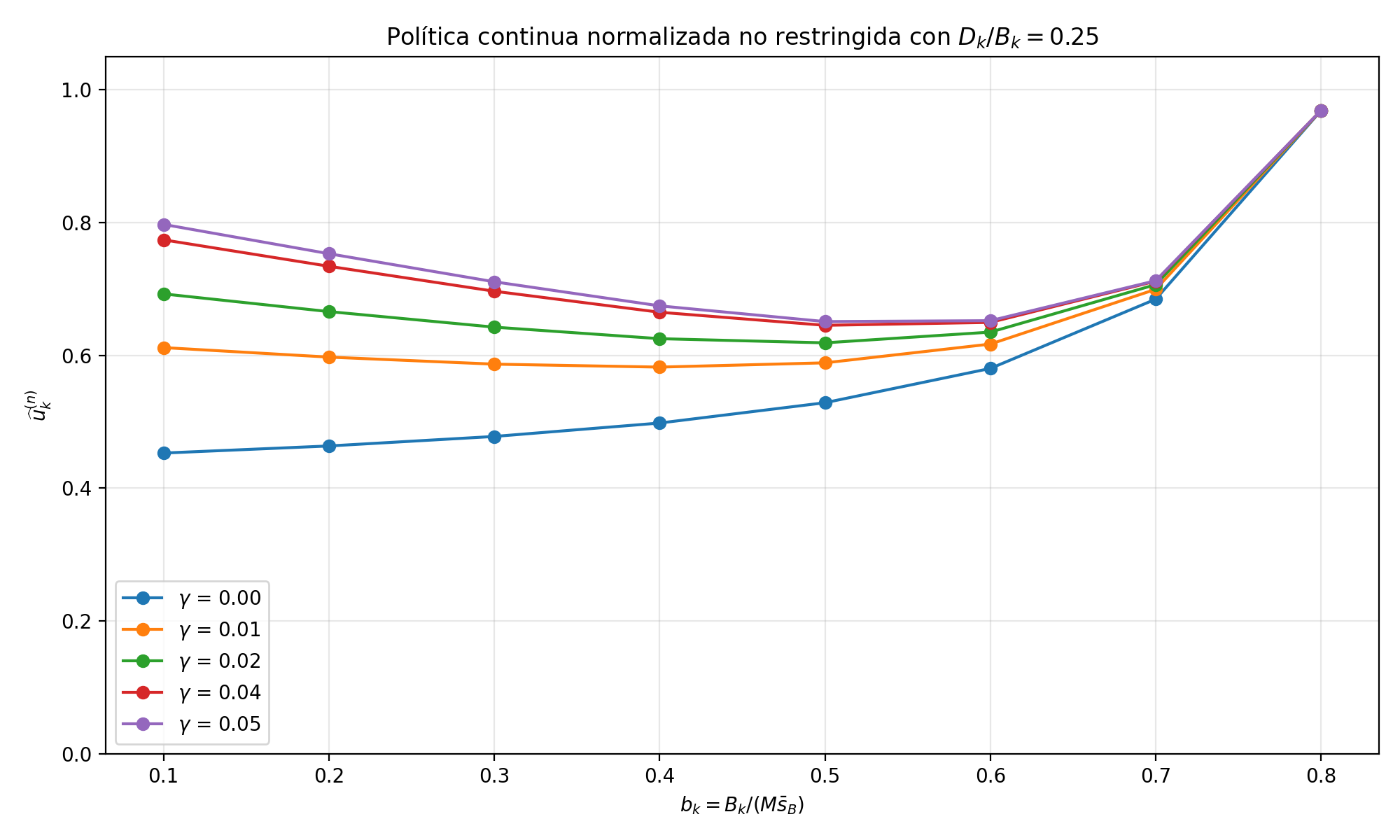}
    \caption{Sensitivity to variations in $\gamma$}
    \label{fig:model_03}
\end{figure}

\subsubsection{Sensitivity to variations in $\theta$}

The model is used to evaluate the sensitivity to variations in the $\theta$ parameter.

\begin{figure}[H]
    \centering
    \includegraphics[width=0.95\linewidth]{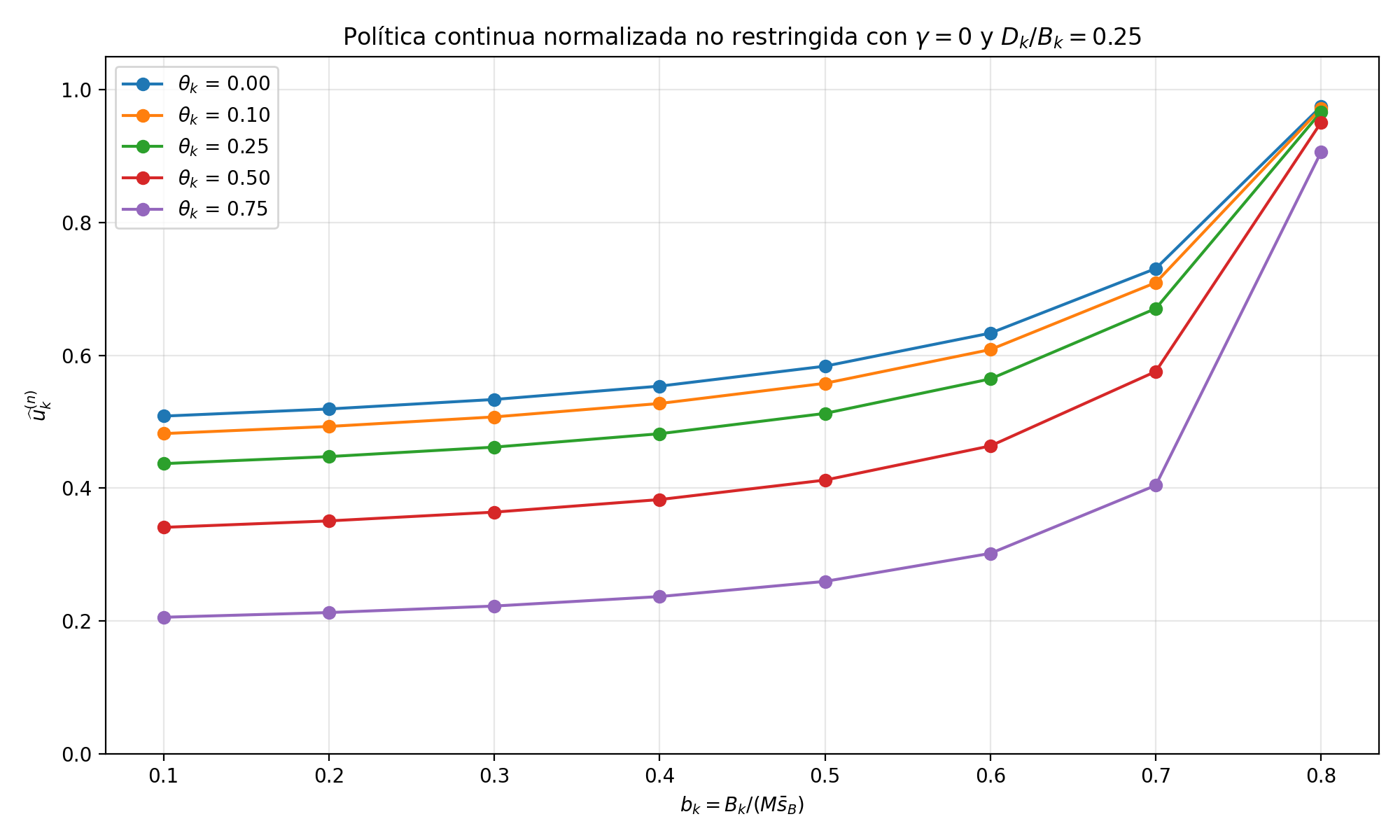}
    \caption{Sensitivity to variations in $\theta_k$ for $\gamma=0$}
    \label{fig:model_04}
\end{figure}

\subsubsection{Sensitivity to variations in $\beta$}

The model is used to evaluate variations in the $\beta$ parameter.

\begin{figure}[H]
    \centering
    \includegraphics[width=0.95\linewidth]{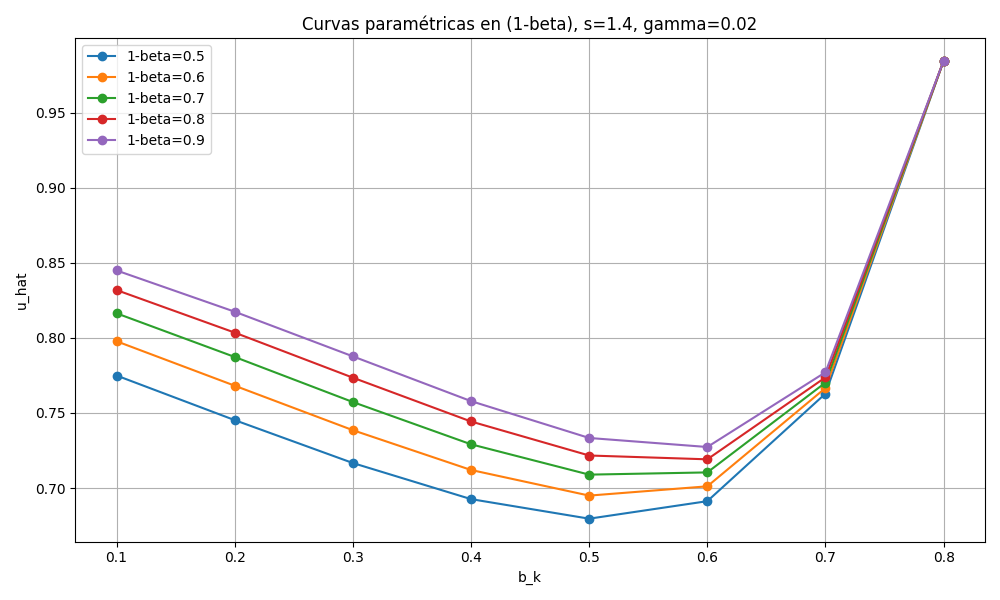}
    \caption{Sensitivity to variations in $\beta$}
    \label{fig:model_05}
\end{figure}

\subsubsection{Sensitivity to variations in $s$}

The model is used to explore variations in the s parameter.

\begin{figure}[H]
    \centering
    \includegraphics[width=0.95\linewidth]{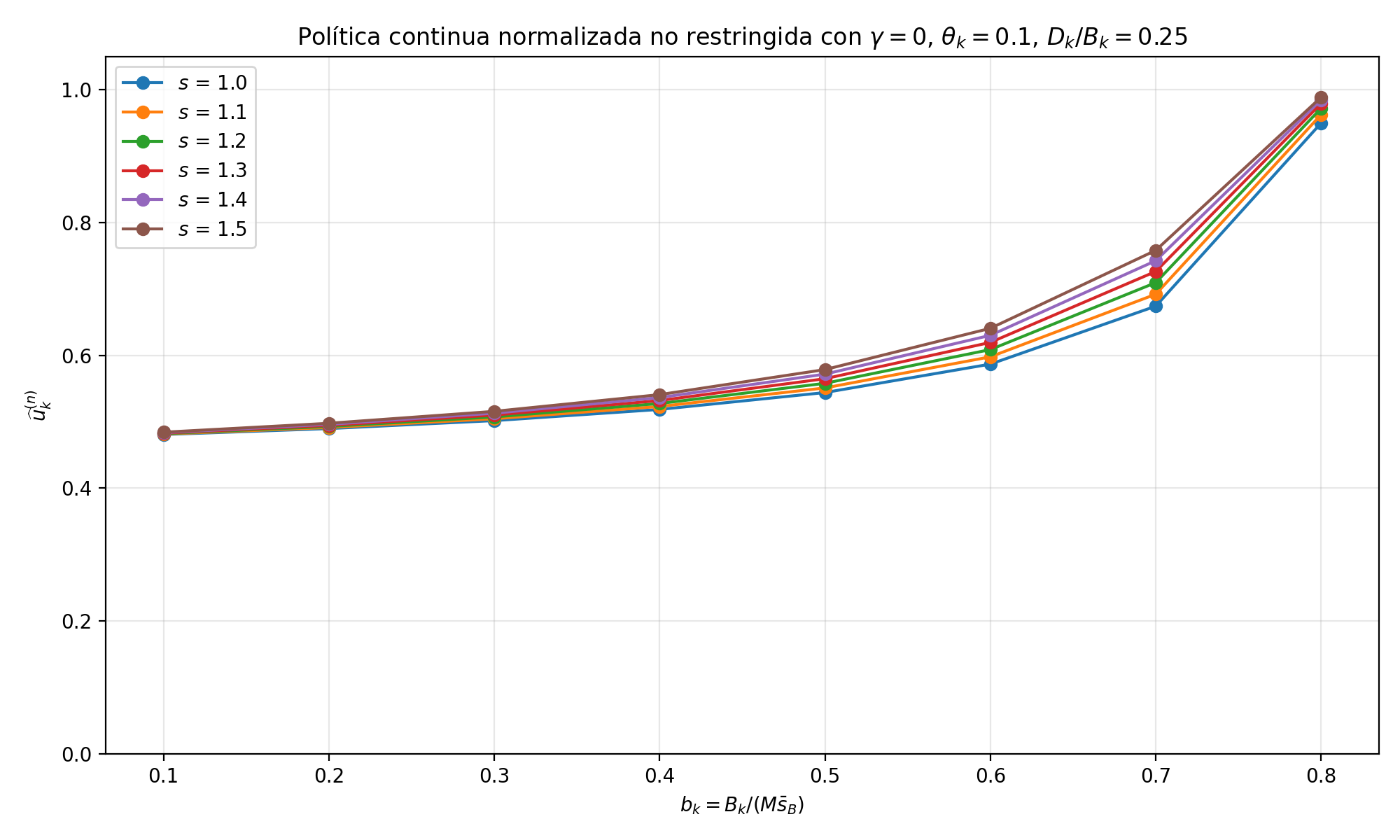}
    \caption{Sensitivity to variations in $s$}
    \label{fig:model_06}
\end{figure}

\subsubsection{Monte Carlo simulation}

A Monte Carlo simulation is performed for a ratio $D_k/B_k=0.5$ using $s_D\approx s_B$ and uniform distributions for all remaining parameters between zero and the maximum estimated values in real situations. Each organization might reproduce the analysis using their own baseline of metrics. The box-plot result is shown for simulation runs at each defined $B_k/M$ ratio.

\begin{figure}[H]
    \centering
    \includegraphics[width=0.95\linewidth]{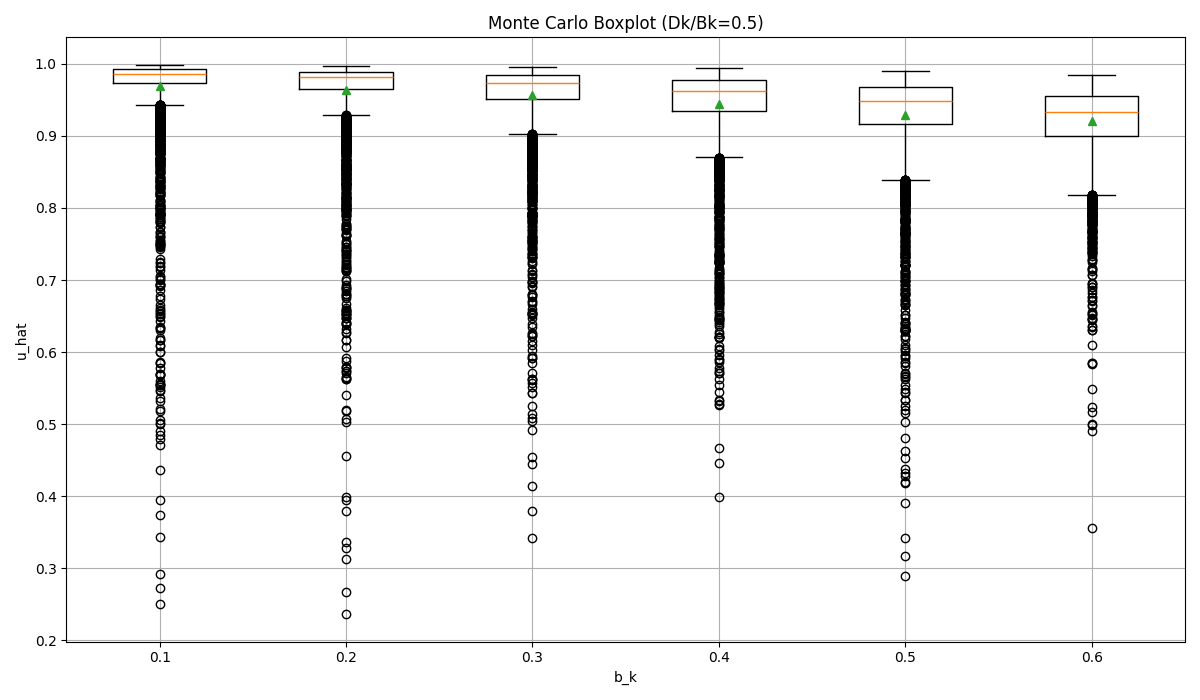}
    \caption{Monte Carlo run}
    \label{fig:model_07}
\end{figure}

\section{Model Limitations}

The model developed in this work should be interpreted as a preliminary and aggregate formulation to support the economic decision between developing new functionality and remediating involuntary technical debt. Its usefulness depends on a set of assumptions that should be made explicit both to delimit its validity domain and to guide subsequent extensions.

First, the model assumes a certain degree of organizational homogeneity and stability. The parameters $\alpha$, $\beta$, and $\gamma$ respectively represent the debt-generation rate associated with new development, the inefficiency of remediation, and the sensitivity of velocity to the stock of technical debt. For these parameters to have operational meaning, it is assumed that the organization has sufficient process maturity to obtain historically stable, and eventually statistically capable, measurements of these magnitudes. In organizations with highly unstable processes, frequent team changes, abrupt domain variations, significant turnover, or absence of systematic measurement, these parameters may fluctuate in such a way that the model loses descriptive or predictive power. In such cases, $\alpha$, $\beta$, and $\gamma$ should be treated as random variables, uncertainty ranges, or context-dependent functions rather than as relatively constant coefficients.
Additionally, technical debt should be interpreted as a socio-technical phenomenon and not exclusively as an internal property of code. Technical dependencies among components induce coordination needs among people and teams, and the lack of congruence between technical dependencies and communication structures may affect development productivity \cite{cataldo2008sociotechnical}. Complementarily, the literature on social debt shows that organizational decisions, deficient coordination, barriers among teams, or suboptimal development communities may generate future costs analogous to those of technical debt \cite{tamburri2013socialdebt,tamburri2015socialdebt}. Consequently, the parameters $\alpha$, $\beta$, and $\gamma$ summarize not only technical properties of the product, but also organizational and socio-technical conditions of the environment in which the product evolves.

Second, the model assumes that organizational participants are able and willing to optimize intertemporal value. This hypothesis is analogous to the rationality assumption used in finance analysis, where agents compare benefits and costs distributed over time under some criterion of present value, opportunity cost, or expected utility. From this perspective, the decision to allocate a fraction $u_k$ of sprint capacity to remediation is interpreted as an intertemporal economic decision: sacrificing part of the immediate value associated with new development may be justified if debt reduction sufficiently improves future delivery capacity. However, in real contexts organizations do not always behave according to this assumption. Political constraints, short-term incentives, commercial pressure, power structures, local performance metrics, or information asymmetries may induce decisions that do not maximize the intertemporal value of the product. In such situations, the model should not be interpreted as a behavioral description of what the organization will actually do, but as a normative reference against which observed decisions can be compared.

Third, the model assumes that the economic value of each story or backlog item is approximately separable from the value of the others, or that coupling among stories is weak. This condition makes it possible to represent the backlog through a decreasing marginal-value structure and to analyze value capture as a function of cumulative progress over the remaining stock. Nevertheless, in real products there may be strong interactions among stories: complementarities, technical dependencies, platform effects, enabling functionalities, incompatibilities, economies of scope, or minimum value packages. In such cases, the value of a story may depend on other stories having been previously developed or being developed jointly. If such couplings are significant, the aggregate value function used by the model may under- or overestimate the economic benefit of certain prioritization decisions.

These limitations do not invalidate the proposed structure, but they do specify its scope. The model is more appropriate when the organization has reasonably stable historical data, when there is willingness to evaluate decisions under a logic of intertemporal value, and when the backlog can be approximated as a set of items whose marginal value is tractable in aggregate terms. When these assumptions are not met, the model may be extended by incorporating parametric uncertainty, state-dependent value functions, explicit organizational constraints, or dependency structures among stories. Consequently, the formulation should be understood as a macroscopic and economic basis for technical-debt management, not as an exhaustive representation of all organizational, political, and architectural factors present in real agile projects.

\section{Conclusions}

This paper has presented a preliminary model for managing involuntary technical debt in agile software development environments. The central contribution is the formulation of technical debt, backlog, velocity, remediation effort, and economic value as a single dynamic decision problem at sprint level. Rather than treating debt merely as an accounting metaphor or as an undifferentiated defect list, the model distinguishes involuntary technical debt from ordinary \textit{defect backlog} and from rework, and represents it as initiated but unfinished functional work that competes with new functionality for limited team capacity.

The proposed formulation shows that the intuitive policy of remediating debt as fast as possible is not necessarily economically optimal. Although immediate remediation may reduce future friction, it can also delay the realization of high-value backlog items. The model therefore introduces the decision variable $u_k$, representing the fraction of sprint capacity assigned to debt remediation, and derives a dynamic trade-off between short-term value capture and long-term productivity recovery. This provides a formal basis for comparing debt-first, feature-first, fixed-allocation, threshold-based, proportional, target-velocity, and cost-based policies within a common economic framework.

A second contribution is the explicit incorporation of backlog value. By modeling backlog items through a decreasing marginal-value structure, the paper connects agile prioritization with intertemporal economic reasoning. This makes it possible to evaluate remediation not only as a technical activity but also as an investment decision: debt reduction is justified when the productivity gained in future sprints compensates for the immediate value deferred by not developing new functionality. In this interpretation, the degradation of velocity induced by debt acts as a form of technical-debt interest.

The model was also extended beyond the continuous aggregate case. The discrete formulation recognizes that real backlog and debt items are indivisible, heterogeneous, and may differ in size, value, and remediation effort. This extension makes the model closer to practical sprint planning, where decisions are made over concrete stories, defects, refactorings, or remediation tasks rather than over perfectly divisible stocks. The comparison among rounding, floor-based, and knapsack-type policies shows that discreteness can materially affect the feasible allocation of capacity and should not be ignored in operational use.

The exploratory validation through sensitivity analysis and Monte Carlo simulation supports the internal coherence of the model. The simulated behavior is consistent with the expected economic intuition: the optimal remediation fraction responds to the relative magnitude of debt, backlog, productivity degradation, value capture, and remediation efficiency. The analysis also shows that the model is especially useful as a reasoning tool for understanding policy sensitivity, parameter uncertainty, and boundary conditions, rather than as a universal closed-form prescription.

At the same time, the paper has made explicit the main limitations of the approach. The formulation is macroscopic: it summarizes heterogeneous and localized technical phenomena through aggregate variables such as $D_k$, $B_k$, $V_k$, and parameters such as $\alpha$, $\beta$, and $\gamma$. It assumes that the organization has sufficient process maturity to estimate these parameters with some degree of statistical stability. It also assumes that participants are willing and able to reason in terms of intertemporal value, and that backlog value is reasonably well prioritized and only weakly coupled across stories. In real organizations, political prioritization, unstable processes, architectural hotspots, socio-technical dependencies, and strong interactions among backlog items may require richer extensions.

Future work should therefore proceed in three directions. First, the model should be empirically calibrated using historical data from agile projects, including velocity, defect arrival, remediation effort, backlog value, and technical-debt indicators. Second, the aggregate formulation should be extended to account for localized debt hotspots, architectural dependencies, and socio-technical coordination structures. Third, the decision policy should be evaluated under uncertainty, using stochastic optimization or simulation-based approaches that treat $\alpha$, $\beta$, $\gamma$, and backlog value as uncertain and context-dependent quantities.

Overall, the paper provides a formal starting point for treating involuntary technical debt as an economic and dynamic management problem. Its main value is not to eliminate managerial judgment, but to make explicit the assumptions, trade-offs, and opportunity costs involved in deciding how much agile capacity should be spent on remediation and how much should be preserved for new value delivery.


\appendix
\section*{Appendix A: Estimation of Value-Relationship Parameters}
\label{appendix:A}
\setcounter{equation}{0}
\renewcommand{\theequation}{A.\arabic{equation}}
The model assumes a value relationship in a prioritized backlog approximately characterized by a Pareto/Zipf distribution. Let a sequence of values ordered in decreasing form be:
\begin{equation}
\{x_1, x_2, \dots, x_n\}
\end{equation}
where each \(x_k\) represents the value associated with the item in position \(k\). It is assumed that the data follow a Pareto/Zipf-type law:
\begin{equation}
x_k = \frac{C}{k^s}
\end{equation}
where \(C > 0\) is a scale parameter and \(s > 0\) is the exponent of the law. Applying natural logarithms to both sides:
\begin{equation}
\ln x_k = \ln C - s \ln k
\end{equation}
Defining:
\begin{equation}
y_k = \ln x_k
\end{equation}
\begin{equation}
z_k = \ln k
\end{equation}
the linear model is obtained:
\begin{equation}
y_k = a + b z_k + \varepsilon_k
\end{equation}
where:
\begin{equation}
a = \ln C
\end{equation}
\begin{equation}
b = -s
\end{equation}

\subsubsection{Estimation by linear regression}

The estimation of the parameters is obtained by ordinary least squares:
\begin{equation}
\hat{b} =
\frac{
\sum_{k=1}^n (z_k - \bar{z})(y_k - \bar{y})
}{
\sum_{k=1}^n (z_k - \bar{z})^2
}
\end{equation}
\begin{equation}
\hat{a} = \bar{y} - \hat{b}\bar{z}
\end{equation}
where:
\begin{equation}
\bar{y} = \frac{1}{n}\sum_{k=1}^n y_k
\end{equation}
\begin{equation}
\bar{z} = \frac{1}{n}\sum_{k=1}^n z_k
\end{equation}
The original parameters are recovered as:
\begin{equation}
\hat{\alpha} = -\hat{b}
\end{equation}
\begin{equation}
\hat{C} = e^{\hat{a}}
\end{equation}
The model can be written in matrix form as:
\begin{equation}
\mathbf{y} =
\begin{bmatrix}
\ln x_1 \\
\ln x_2 \\
\vdots \\
\ln x_n
\end{bmatrix}
\end{equation}
\begin{equation}
\mathbf{X} =
\begin{bmatrix}
1 & \ln 1 \\
1 & \ln 2 \\
\vdots & \vdots \\
1 & \ln n
\end{bmatrix}
\end{equation}
\begin{equation}
\hat{\beta} =
\begin{bmatrix}
\hat{a} \\
\hat{b}
\end{bmatrix}
=
(\mathbf{X}^T \mathbf{X})^{-1}\mathbf{X}^T \mathbf{y}
\end{equation}

\subsubsection{Bias of the logarithmic estimator}

It should be noted that:
\begin{equation}
E[\ln x] \neq \ln E[x]
\end{equation}
which implies that log-log regression introduces bias in finite samples.

\subsubsection{Alternative estimation: maximum likelihood}

Assuming a continuous Pareto distribution:
\begin{equation}
f(x) = d x_{min}^s x^{-(s+1)}
\end{equation}
the maximum-likelihood estimator is:
\begin{equation}
\hat{\alpha}
=
\left[
\frac{1}{n}
\sum_{i=1}^n
\ln \frac{x_i}{x_{min}}
\right]^{-1}
\end{equation}
\subsubsection{Extension: truncated Zipf law}
To improve the fit in real data, the following can be used:
\begin{equation}
x_k = \frac{C}{(k + \delta)^s}
\end{equation}
with \(\delta > 0\). Applying logarithms:
\begin{equation}
\ln x_k = \ln C - s \ln(k+\delta)
\end{equation}
This model is not linear in the parameters and requires nonlinear least-squares estimation.

\subsubsection{Interpretation in the technical-debt model}

The parameter \(\alpha\) characterizes value concentration:

\begin{itemize}
\item high $s > 0$ implies strong value concentration in a few items;
\item \(s \approx 1\) corresponds to classic Zipf-like behavior;
\item \(s < 1\) indicates a heavy tail and a more homogeneous distribution.
\end{itemize}

This directly impacts the optimal effort allocation in the main model.

\subsubsection{Relationship with packing efficiency}

The heterogeneity induced by \(\alpha\) affects sprint efficiency:
\begin{equation}
\eta_k = 1 - \frac{W_k}{V_k}
\end{equation}
where \(W_k\) represents unused capacity due to indivisibilities. In particular:
\begin{equation}
s \uparrow \quad \Rightarrow \quad \eta_k \downarrow
\end{equation}
which connects the distribution of value with the operational efficiency of the sprint.

\section*{Appendix B: Empirical Estimation of the Degradation Parameter $\gamma$}
\label{appendix:B}

\setcounter{equation}{0}
\renewcommand{\theequation}{B.\arabic{equation}}

The parameter $\gamma$ models the effect of technical debt on the degradation of the system's value or productivity. Its empirical estimation is essential to calibrate the model.

In the main model, the effective value of the system is expressed as:
\begin{equation}
V_k = V_0 e^{-\gamma D_k}
\end{equation}
or in its approximate rational form:
\begin{equation}
V_k = \frac{V_0}{1+\gamma D_k}
\end{equation}
Both formulations imply that $\gamma$ measures the sensitivity of velocity with respect to accumulated debt.

\subsection{Required data}

To estimate $\gamma$, historical observations are needed:
\begin{equation}
\{(D_k, V_k)\}_{k=1}^n
\end{equation}
where:

\begin{itemize}
\item $D_k$: technical-debt level in sprint $k$;
\item $V_k$: observed system velocity in sprint $k$.
\end{itemize}

Velocity ($V_k$) is understood to be expressed as the number of effort units (usually story points) per sprint, although alternative measures may also be used, such as:

\begin{itemize}
\item effective team velocity;
\item throughput of completed stories;
\item delivered economic value (story points weighted by value);
\item adjusted productivity metrics.
\end{itemize}

\subsubsection{Estimation under the exponential model}

Starting from:
\begin{equation}
V_k = V_0 e^{-\gamma D_k}
\end{equation}
Applying logarithms:
\begin{equation}
\ln V_k = \ln V_0 - \gamma D_k
\end{equation}
Defining:
\begin{equation}
y_k = \ln V_k
\end{equation}
\begin{equation}
x_k = D_k
\end{equation}
the linear model is obtained:
\begin{equation}
y_k = a + b x_k + \varepsilon_k
\end{equation}
where:
\begin{equation}
a = \ln V_0
\end{equation}
\begin{equation}
b = -\gamma
\end{equation}

The least-squares estimation is:
\begin{equation}
\hat{b} =
\frac{
\sum_{k=1}^n (x_k - \bar{x})(y_k - \bar{y})
}{
\sum_{k=1}^n (x_k - \bar{x})^2
}
\end{equation}
\begin{equation}
\hat{a} = \bar{y} - \hat{b}\bar{x}
\end{equation}

therefore:
\begin{equation}
\hat{\gamma} = -\hat{b}
\end{equation}
\subsubsection{Estimation under the rational model}

Considering:
\begin{equation}
V_k = \frac{V_0}{1+\gamma D_k}
\end{equation}
it can be rewritten as:
\begin{equation}
\frac{1}{V_k} = \frac{1}{V_0} + \frac{\gamma}{V_0} D_k
\end{equation}
Defining:
\begin{equation}
y_k = \frac{1}{V_k}
\end{equation}
\begin{equation}
x_k = D_k
\end{equation}
we obtain:
\begin{equation}
y_k = a + b x_k
\end{equation}
where:
\begin{equation}
a = \frac{1}{V_0}
\end{equation}
\begin{equation}
b = \frac{\gamma}{V_0}
\end{equation}
Then:
\begin{equation}
\hat{\gamma} = \frac{\hat{b}}{\hat{a}}
\end{equation}

\subsubsection{Direct nonlinear estimation}

Alternatively, $\gamma$ may be estimated by solving:
\begin{equation}
\min_{\gamma, V_0}
\sum_{k=1}^n
\left(
V_k - V_0 e^{-\gamma D_k}
\right)^2
\end{equation}

or:

\begin{equation}
\min_{\gamma, V_0}
\sum_{k=1}^n
\left(
V_k - \frac{V_0}{1+\gamma D_k}
\right)^2
\end{equation}

This approach avoids transformations and reduces bias, but requires iterative methods.

\subsection{Parameter interpretation}

The parameter $\gamma$ can be interpreted as:
\begin{equation}
\gamma = -\frac{1}{V_k} \frac{\partial V_k}{\partial D_k}
\end{equation}
which represents the marginal sensitivity of velocity to debt. Typical values:
\begin{itemize}
\item high $\gamma$: strong degradation due to debt;
\item low $\gamma$: resilient system.
\end{itemize}

\subsubsection{Consistency with the discrete model}

The estimation of $\gamma$ influences the optimal policy $u_k^\star$ of the main model, because it affects the marginal valuation of reducing debt relative to producing backlog.
In particular:

\begin{equation}
\gamma \uparrow \quad \Rightarrow \quad u_k^\star \uparrow
\end{equation}
which implies greater optimal allocation toward technical-debt reduction.

\section*{Appendix C: Characterization and Estimation of the Economic Parameter $\lambda$}
\label{appendix:C}

\setcounter{equation}{0}
\renewcommand{\theequation}{C.\arabic{equation}}

Let $\nu_k$ be the economic value generated by the system in sprint $k$, and $V_k$ the team velocity measured in story points delivered in that sprint.

The parameter $\lambda$ is defined as the marginal economic value of one additional unit of future capacity:
\begin{equation}
\lambda = \frac{\partial \nu_k}{\partial V_k}
\end{equation}
This parameter represents the shadow price of the productive capacity of the system.

\subsubsection{Relationship with productivity improvement}

If a reduction in technical debt produces an increase in future velocity $\Delta V_k$, the increase in value can be approximated by:
\begin{equation}
\Delta \nu_k \approx \lambda \, \Delta V_k
\end{equation}
which justifies the use of $\lambda$ in equations (84) and (85) of the main model.

\subsubsection{Empirical model}

To estimate $\lambda$, it is proposed to model the relationship between economic value and velocity:
\begin{equation}
\nu_k = \nu_0 + \lambda V_k + \varepsilon_k
\end{equation}
where:

\begin{itemize}
\item $\nu_k$: economic value generated in sprint $k$;
\item $V_k$: velocity (delivered story points);
\item $\nu_0$: base value independent of velocity;
\item $\varepsilon_k$: error term.
\end{itemize}

\subsubsection{Estimation by linear regression}

Defining:
\begin{equation}
y_k = \nu_k
\end{equation}
\begin{equation}
x_k = V_k
\end{equation}
we obtain:
\begin{equation}
y_k = a + b x_k + \varepsilon_k
\end{equation}
where:
\begin{equation}
a = \nu_0
\end{equation}
\begin{equation}
b = \lambda
\end{equation}

The least-squares estimator is:
\begin{equation}
\hat{\lambda} =
\frac{
\sum_{k=1}^n (x_k - \bar{x})(y_k - \bar{y})
}{
\sum_{k=1}^n (x_k - \bar{x})^2
}
\end{equation}
with:
\begin{equation}
\bar{x} = \frac{1}{n}\sum_{k=1}^n V_k
\end{equation}
\begin{equation}
\bar{y} = \frac{1}{n}\sum_{k=1}^n \nu_k
\end{equation}

\subsubsection{Incremental estimation}

To eliminate the base term $\nu_0$, a differences formulation can be used:
\begin{equation}
\Delta \nu_k = \nu_k - \nu_{k-1}
\end{equation}
\begin{equation}
\Delta V_k = V_k - V_{k-1}
\end{equation}
and fit:
\begin{equation}
\Delta \nu_k = \lambda \, \Delta V_k + \varepsilon_k
\end{equation}

\subsubsection{Required data}

To estimate $\lambda$, the following dataset is required:
\begin{equation}
\{(V_k, \nu_k)\}_{k=1}^n
\end{equation}
where:

\begin{itemize}
\item $V_k$: sprint velocity (completed story points);
\item $\nu_k$: generated economic value.
\end{itemize}

The value $\nu_k$ can be approximated through:

\begin{itemize}
\item revenue generated in the sprint;
\item business value assigned to completed stories;
\item story points weighted by relative value;
\item economic proxies derived from operational metrics.
\end{itemize}

\subsubsection{Economic interpretation}
The parameter $\lambda$ represents:
\begin{equation}
\lambda > 0
\end{equation}
and its magnitude indicates the marginal value of capacity:
\begin{itemize}
\item high $\lambda$: strong return from increasing velocity;
\item low $\lambda$: decreasing returns.
\end{itemize}

\subsubsection{Nonlinear extension}

If the relationship is not linear, one may consider:
\begin{equation}
\nu_k = f(V_k)
\end{equation}
and define:
\begin{equation}
\lambda = \frac{d f(V_k)}{d V_k}
\end{equation}
In this case, linear regression estimates an average value of $\lambda$ over the observed range.

\subsubsection{Consistency with the decision model}

The value of $\lambda$ determines the economic benefit of increasing future velocity, and therefore directly influences the optimal decision of effort allocation between backlog and technical debt. In particular:

\begin{equation}
\lambda \uparrow \quad \Rightarrow \quad u_k^\star \uparrow
\end{equation}
which implies a greater incentive to invest in debt reduction in order to improve future capacity.

\end{document}